\documentclass[fleqn,a4paper,12pt]{article}
\usepackage{amssymb}
\usepackage{makeidx}
\usepackage{amsmath}
\usepackage{graphicx}
\usepackage{lscape}
\usepackage{a4}
\usepackage{epsfig}
\renewcommand{\bar}{\overline} \newcommand{\newc}{\newcommand}
\newc{\beq}{\begin{equation}} \newc{\eeq}{\end{equation}}
\newc{\bea}{\begin{array}} \newc{\eea}{\end{array}}
\newc{\ri}{{\mathrm i}}
\newc{\bW}{{\mathbf W}}
\newc{\bR}{{\mathbf R}}
\newc{\bN}{{\mathbf N}}
\newc{\Psibar}{\overline\Psi}
\newc{\w}{{\bf w}}
\newc{\E}{{\mathbf{E}}}
\newc{\bp}{{\bf p}}
\newc{\ta}{\tilde a}
\newc{\bV}{{\bf V}}
\newc{\bfV}{{\bf V}}
\newc{\bfG}{{\bf G}}
\newc{\bx}{{\bf x}}
\newc{\bu}{{\bf u}}
\newc{\bP}{{\bf P}}
\newc{\bJ}{{\bf J}}
\newc{\bK}{{\bf K}}
\newc{\pd}{{\partial}}
\newc{\ti}{{\times}}
\newc{\bA}{{\bf A}}
\newc{\bE}{{\bf E}}
\newc{\bfn}{{\bf\nabla}}
\newc{\ho}{\hookrightarrow}
\newc{\ra}{\rightarrow}
\newc{\bv}{{\bf v}}
\newc{\bb}{{\bf b}}
\newc{\bc}{{\bf c}}
\newc{\bd}{{\bf d}}
\newc{\tbb}{\tilde{\bf b}}
\newc{\tbc}{\tilde{\bf c}}
\newc{\tbd}{\tilde{\bf d}}
\newc{\bz}{{\bf 0}}
\newc{\bun}{{\bf 1}}
\newc{\bL}{{\bf L}}
\newc{\bS}{{\bf S}}
\newc{\bB}{{\bf B}}
\newc{\br}{{\bf r}}
\newc{\sig}{{\mathbf\sigma}}
\newc{\eg}{{\it e.g.\ }}
\newc{\bpi}{{\mathbf\pi}}
\newc{\ie}{{\it i.e.\ }}
\newc{\etal}{{\it et al}}
\newc{\AM}{{Ann. Math.}}
\newc{\NCB}{Nuov. Cim. {\bf B}}
\newc{\JPA}{J. Phys. A: Math. Gen.}
\newc{\CMP}{Comm. Math. Phys.}
\newc{\PRD}{Phys. Rev. D}
\newc{\TMP}{Theor. Math. Phys.}
\newc{\JPG}{J. Phys. G: Nucl. Part. Phys.}
\catcode`@=11 \long
\def\@caption#1[#2]#3{\par\addcontentsline{\csname
ext@#1\endcsname}{#1} {\protect\numberline{\csname
the#1\endcsname}{\ignorespaces #2}} \begingroup \small
\@parboxrestore \@makecaption{\csname fnum@#1\endcsname}
{\ignorespaces #3}\par \endgroup} \catcode`@=12

\def\JPA#1#2#3#4{#2 #1 {\em J. Phys. A: Math. Gen.} {\bf #3} #4}

\def\AM#1#2#3#4{#2 #1 {\em Ann. Math.} {\bf #3} #4}

\def\NCB#1#2#3#4{#2 #1 {\em Nuov. Cim.} {\bf #3 B} #4}

\def\PRD#1#2#3#4{#2 #1 {\em Phys. Rev.} {\bf D #3} #4}

\def\IJTP#1#2#3#4{#2 #1 {\em Int. J. Theor. Phys.} {\bf #3} #4}
\def\CMP#1#2#3#4{#2 #1 {\em Comm. Math. Phys.} {\bf #3} #4}

\def\TMP#1#2#3#4{#2 #1 {\em Theor. Math. Phys.} {\bf #3} #4}

\def\RMP#1#2#3#4{#2 #1 {\em Rep. Math. Phys.} {\bf #3} #4}
\def\TMP#1#2#3#4{#2 #1 {\em Theor. Math. Phys.} {\bf #3} #4}

\def\RMP#1#2#3#4{#2 #1 {\em Rev. Mod. Phys.} {\bf #3} #4}

\def\JPG#1#2#3#4{#2 #1 {\em J. Phys. G: Nucl. Part. Phys.} {\bf
#3} #4}

\begin{document} \begin{titlepage} \vskip 2cm
\begin{center} {\bf \Large Galilei-invariant equations for massive
fields} \vskip 3cm {\bf
J. Niederle$^a$, \\ and A.G. Nikitin$^b$ } \vskip 5pt {\sl
$^a$Institute of Physics of the Academy of Sciences of the Czech
Republic,\\ Na Slovance 2, 18 221 Prague, Czech Republic} \footnote{E-mail: {\tt
niederle@fzu.cz} } \vskip 2pt
{\sl $^b$Institute of Mathematics, National Academy of Sciences of
Ukraine,\\ 3 Tereshchenkivs'ka Street, Kyiv-4, Ukraine, 01601
\footnote{E-mail: {\tt nikitin@imath.kiev.ua} }\\}
\end{center}
\vskip .5cm \rm

\begin{abstract} Galilei-invariant equations for massive fields with various spins have been
found and classified. They have been derived directly, i.e., by using
requirement of the Galilei invariance and various facts on representations
of the Galilei group deduced in the paper written by de Montigny M, Niederle
J and Nikitin A G, J. Phys. A \textbf{39}, 1-21, 2006. A completed list of
non-equivalent Galilei-invariant wave equations for vector and scalar fields
is presented.
It shows two things. First  that the collection of such equations is very
broad and describes many physically consistent systems.
In particular it is possible to describe spin-orbit and Darwin couplings
in frames of Galilei-invariant approach.
Second, these Galilei-invariant equations can be obtained either
via contraction of known
relativistic equations
 or
 via contractions of quite new relativistic wave equations.
\end{abstract}

\end{titlepage}

\setcounter{footnote}{0} \setcounter{page}{1}
\setcounter{section}{0}


\section{\label{S1}Introduction}

It is well known that the Galilei group $G(1,3)$ and its representations
play the same role in non-relativistic physics as the Poincar\'{e} group $%
P(1,3)$ and its representations do in the relativistic case. In fact the
Galilei group and its representations form a group-theoretical basis of
classical mechanics and electrodynamics. They replace the Poincar\'{e} group
and its representations whenever velocities of bodies are much smaller then
the speed of light in vacuum. On the other hand, structure of subgroups of
the Galilei group and of its representations are in many respects more
complex than those of the Poincar\'{e} group and therefore it is not perhaps
so surprising that representations of the Poincar\'{e} group were described
in \cite{Wigner3} almost 15 years earlier than representations of the
Galilei group \cite{bargman} in spite of the fact that the relativity
principle of classical physics was formulated by Galilei in 1632, i.e.,
about three centuries prior to that of relativistic physics by Einstein.

An excellent review of representations of the Galilei group was written by L%
\'{e}vy-Leblond \cite{levyleblond}. It appears that the Galilei group, as
distinct from the Poincar\'{e} group, has besides ordinary representations
also projective ones (see \cite{bargman} and \cite{Wigner} respectively).
However its subgroup -- the homogeneous Galilei group $HG(1,3)$ which plays
in non-relativistic physics the role of the Lorentz group in the
relativistic case -- has a more complex structure so that its
finite-dimensional indecomposable representations are not classifiable (for details see \cite{Marc}). And they are the representations
which play a key role in description of physical systems satisfying the
Galilei relativity principle!

An important class of indecomposable finite-dimensional representations of
the group $HG(1,3)$ was found and completely classified in \cite{Marc}. It
contains all representations of the homogeneous Galilei group which, when
restricted to its rotation subgroup, decompose to spin 0, 1/2 and 1
representations. It was explained in \cite{Marc} and \cite{NN2} how these
representations can be obtained from those of the Lorentz group by means of
the In\"{o}n\"{u}-Wigner contractions.

The representations classified in \cite{Marc} are interesting from
both physical and mathematical points of view. Physically, the
related spin values exhaust all ones observed experimentally for
stable particles. Mathematically, the results of paper \cite{Marc}
are unimprovable in the sense that the problem of classification of
the representations with other spin content is unsolvable in
general. This gives us undreamt possibilities to describe various
non-relativistic (quantum-mechanical and field-theoretical) systems
of interacting particles and fields with spins 0, 1/2 and 1. For
instance, it was possible to find the most general Pauli interaction
of the Galilean spin--1/2 particles with an external electromagnetic
field, see \cite{Marc}.

Starting with indecomposable representations of the group $HG(1,3)$
found in \cite{Marc} an extended class of linear and non-linear
equations for Galilean massless fields has been found in our
previous paper \cite{NN3}. In particular, Galilean analogues of the
Born-Infeld and Maxwell-Shern-Simon systems were discussed there.
Moreover, all possible Galilei-invariant  systems of first order
equations for massless vector and scalar fields had been classified
in \cite{NN3}.

In the present paper we continue and complete the research started
in \cite{Marc}-\cite{NN3}. Namely, we study first vector and spinor
representations of the homogeneous Galilei group in detail and then
using them we construct various wave equations for {\it massive}
particles with spin 0, 1/2, 1 and 3/2. In this way a completed list
of systems of Galilei invariant first order partial differential
equations is found which describe particles with spin $s<3/2$.

 There are well--developed methods for deriving relativistic wave equations
invariant w.r.t. the Poincar\'{e} group\footnote{These methods have
had a long a tangled history. It began with Schr\"odinger's proposal
of a relativistic equation in 1926, Dirac's equation for massive
spin 1/2 particle in 1928 and Majorana's pioneering study in 1932
and has always been a subject of interest of many scientists
(Bhabha, Harish-Chandra, Wild, Fierz, Pauli, Kemmer, Duffin,
Bargmann, Wigner, Weinberg, Schwinger, and so on, see, e.g.,
\cite{7} and references cited therein)}. Consequently several
systematic approaches to relativistic equations are now available,
the most popular of which are those by Bhabha and Gel'fand and
Yaglom \cite{8}, those by Bruhat \cite{9} and those based on G\aa
rding's technigue \cite{10}. These approaches can be used for
construction of the Galilei-invariant equations as well.

First, we begin with the Bhabha and Gel'fand-Yaglom \cite{8} approach which
is a direct extension of the method yielding the Dirac equation. The
corresponding relativistic wave equations can be written as systems of the
first order partial-differential linear equations of the form:
\begin{equation}
\left( \beta _{\mu }p^{\mu }+\beta _{4}m\right) \Psi (\mathbf{x},t)=0,
\label{equ}
\end{equation}%
where $p^{0}=i\frac{\partial }{\partial x_{0}},\ p^{a}=i\frac{\partial }{%
\partial x_{a}}\ (a=1,2,3),$ and $\beta _{\mu }\ (\mu =0,1,2,3)$ and $\beta
_{4}$ are square matrices restricted by the condition of the Poincar\'{e}
invariance. Notice that in the relativistic approach matrix $\beta _{4}$ is
usually assumed to be proportional to a unit matrix.

The theory of the Poincar\'{e}-invariant equations (\ref{equ}) is clearly
explained in detail for instance in the Gel'fand--Minlos--Shapiro book \cite%
{gelfand}.

The other approaches make use of tensor calculus, and the associated
equations have the form of covariant vectors or tensors (see, e.g., \cite%
{corson}). A popular example is the first order Proca equation
\cite{proca},
\begin{equation}
\begin{array}{l}
p^{\mu }\Psi ^{\nu }-p^{\nu }\Psi ^{\mu }=\kappa \Psi ^{\mu \nu }, \\
p_{\nu }\Psi ^{\nu \mu }=\kappa \Psi ^{\mu },\label{001}%
\end{array}%
\end{equation}%
where $\Psi ^{\mu }$ and $\Psi ^{\mu \nu }$ is a four-vector and a
skew-symmetric spinor respectively which transform according to the
representation $D(\frac{1}{2},\frac{1}{2})\oplus D(1,0)\oplus D(0,1)$ of the
Lorentz group. The second order Proca \cite{proca}, Rarita-Schwinger \cite%
{rarita} and Singh--Hagen \cite{sign} equations serve as other examples. Let
us mention that all these relativistic equations violate causality or
predict incorrect values for the gyromagnetic ratio $g$. The
tensor-spinorial equations for particles with an arbitrary half-integer spin
which are not violating causality and admit the right value for $g$ were
discussed in detail in \cite{niederle}.

In the present paper we use both the above mentioned approaches and derive
the Galilei-invariant equations for particles with spins 0, $\frac{1}{2}$, 1
and $\frac{3}{2}$. Moreover we present a complete list of the
Galilei-invariant equations (\ref{equ}) for scalar and vector fields. In
contains equations obtained via contractions of known relativistic equations
as well as new Galilei-invariant equations which are received via
contractions of new relativistic equations.

A special class of Galilei-invariant equations for particles with
arbitrary spins $s$ was described in the significant paper
\cite{Hagen1}. The related equations have the minimal number of
components and predict the same value $g=1/s$ for the gyromagnetic
ratio as it is done by the relativistic wave equations
\cite{Hagen1}.

Some particular results associated with the Galilei-invariant equations (\ref{equ}%
) can also be found in \cite{levyleblond}, \cite{Marc}, \cite{ll1967}
-\cite{ourletter}.
Galilean analogues of the Bargman-Wigner equations are presented in a recent
paper \cite{Mar}. However, these equations became incompatible whenever a
minimal interaction with an external e.m. field had been introduced \cite%
{Mar}.


\section{\label{S2}The Galilei algebra and Galilei--invariant wave equations}

\subsection{\label{S21}Basic definitions}

In this section we shall develop a Galilean version of the Bhabha and
Gel'fand-Yaglom  approach \cite{8} and present a complete list of the
corresponding Galilei-invariant wave equations. Let us note that some of
these equations are already known (see for instance, \cite{levyleblond},
\cite{nikitin} and \cite{ourletter}).

Equation (\ref{equ}) is said to be \textit{invariant} w.r.t. the Galilei
transformations
\begin{equation}  \label{11}
\begin{array}{l}
\mathbf{x}\to \mathbf{x}^{\prime }= \mathbf{R}\mathbf{x} +\mathbf{v}t+%
\mathbf{b}, \ \ \ \ \
t\to t^{\prime }=t+a%
\end{array}%
\end{equation}
where $a,\mathbf{b}$, $\mathbf{v}$ are real parameters and $\mathbf{R}$ is a
rotation matrix, if function $\Psi$ in (\ref{equ}) cotransforms as%
\begin{equation}  \label{cova}
\Psi(\mathbf{x}, t)\to\Psi^{\prime }(\mathbf{x}^{\prime }, t^{\prime}) =\texttt{e}^{\mathrm{i }f(\mathbf{x}, \ t)}T\Psi(\mathbf{x}, t),
\end{equation}
i.e., according to a particular representation of the Galilei group. Here $T$
is a matrix depending on transformation parameters only, $f(\mathbf{x},
t)=m\left(\mathbf{v} \cdot\mathbf{x}+t{v^2}/{2}+c\right)$, $c$ is an
arbitrary constant and $\Psi^{\prime }(\mathbf{x^{\prime }}, t^{\prime })$
satisfies the same equation in prime variables as $\Psi(\mathbf{x}, t)$ in
the initial ones.

The Lie algebra corresponding to representation (\ref{cova}) has the
following generators
\begin{equation}
\begin{array}{l}
P_{0}=p_{0}=\mathrm{i}\frac{\partial }{\partial t},\ \ P_{a}=p_{a}=-\mathrm{i%
}\frac{\partial }{\partial x^{a}},\ M=mI, \\
J_{a}=\varepsilon _{abc}x_{b}p_{c}+S_{a} \\
G_{a}=tp_{a}-mx_{a}+\eta _{a},%
\end{array}
\label{cov}
\end{equation}%
where $S_{a}$ and $\eta _{a}$ are matrices which satisfy the following
commutation relations:
\begin{equation}\bea{l}
\lbrack S_{a},S_{b}]=\mathrm{i}\varepsilon _{abc}S_{c},  \label{e33} \\
{[}\eta _{a},S_{b}{]}=\mathrm{i}\varepsilon _{abc}\eta _{c},\ {[}\eta
_{a},\eta _{b}{]}=0\eea
\end{equation}%
that is, they form a basis of the homogeneous Galilei algebra $hg(1,3)$.

Equation (\ref{equ}) is invariant with respect to the Galilei
transformations (\ref{11}), (\ref{cova}), if their generators (\ref{cov})
transform solutions of (\ref{equ}) into solutions. This requirement together
with the existence of the Galilei-invariant Lagrangian for (\ref{equ})
yields the following conditions on matrices $\beta _{\mu }$ ($\mu =0,1,2,3,$%
) and $\beta _{4}$ \cite{fuschich}:
\begin{equation}
\begin{array}{l}
\label{cond}\eta _{a}^{\dag }\beta _{4}-\beta _{4}\eta _{a}=-i\beta _{a}, \\
\eta _{a}^{\dag }\beta _{b}-\beta _{b}\eta _{a}=-i\delta _{ab}\beta _{0}, \\
\eta _{a}^{\dag }\beta _{0}-\beta _{0}\eta _{a}=0,\ \ a,b=1,2,3.%
\end{array}%
\end{equation}

Moreover, $\beta_0$ and $\beta_4$ must be scalars w.r.t. rotations, i.e.,
they have to commute with $S_a$.

Thus the problem of classification of the Galilei-invariant equations (\ref%
{equ}) is equivalent to find matrices $S_{a},\ \eta _{a},\ \beta _{0},\beta
_{a}$ and $\beta _{4}$ satisfying relations  (\ref{e33}) and (\ref{cond}). Unfortunately, a subproblem of this problem, i.e., a complete
classification of non-equivalent finite--dimensional representations of
algebra (\ref{e33}) appears to be in general unsolvable (that is
a `wild' algebraic problem). However, for two important particular cases,
i.e., for purely spinor and vector-scalar representations, the problem of
finding all finite--dimensional indecomposable representations of the
algebra $hg(1,3)$ is solvable and was completely solved in \cite{Marc}.

\subsection{\label{S22}Spinor fields and the corresponding wave equations}

Let $\tilde s$ be the highest value of spin which appears when
representation of algebra $hg(1,3)$ is reduced to its subalgebra $so(3)$.
Then the corresponding representation space of $hg(1,3)$ is said to be the
space of fields of spin $\tilde s$.

As mentioned in \cite{Marc} there exist only two non-equivalent
indecomposable representations of the algebra $hg(1,3)$ defined on fields of
spin 1/2. One of them, $D_1({\frac12})$, when restricted to the subalgebra $%
so(3)$ remains irreducible while the other one, $D_2(\frac12)$, decomposes
to two irreducible representations $D(1/2)$ of $so(3)$. The corresponding
matrices $S_a$ and $\eta_a$ can be written in the following form:
\begin{equation}  \label{1/2a}
S_a=\frac12\sigma_a, \ \ \eta_a=\mathbf{0} \texttt{\ for }
D_1\left(\frac12\right)
\end{equation}
and
\begin{equation}  \label{02}
S_a=\frac12\left(%
\begin{array}{cc}
\sigma_a & \mathbf{0} \\
\mathbf{0} & \sigma_a%
\end{array}%
\right) ,\ \eta_a=\frac{\mathrm{i}}2\left(%
\begin{array}{cc}
\mathbf{0} & \mathbf{0} \\
\sigma_a & \mathbf{0}%
\end{array}%
\right) \texttt{\ for }D_2\left(\frac12\right).
\end{equation}
Here $\sigma_a$ are the Pauli matrices and $\mathbf{0}$ is a $2\times2$ zero
matrix.

Realization (\ref{1/2a}) with conditions (\ref{cond}) yields equation (\ref%
{equ}) trivial, i.e., with zero $\beta$--matrices.

The elements of the carrier space of representation (\ref{02}) will be
called the Galilean bi-spinors. It can be found in \cite{Marc} how the
Galilean bi-spinors transform w.r.t. the finite transformations from the
Galilei group.

Solutions of relations (\ref{cond}) with $S_a, \eta_a$ given by formulae (%
\ref{02}) can be written as
\begin{equation}
\begin{array}{l}
\beta_0=\left(%
\begin{array}{cc}
\mathtt{I} & \mathbf{0} \\
\mathbf{0} & \mathbf{0}%
\end{array}%
\right), \ \beta_a=\left(%
\begin{array}{cc}
\mathbf{0} & \sigma_a \\
\sigma_a & \mathbf{0}%
\end{array}%
\right),\ \beta_4=\left(%
\begin{array}{cc}
\kappa \mathtt{I} & -\mathrm{i}\omega \mathtt{I} \\
\mathrm{i}\omega \mathtt{I} & 2\mathtt{I}%
\end{array}%
\right), \
a=1,2,3,%
\end{array}
\label{beta}
\end{equation}
where $\mathtt{I}$ and $\mathbf{0}$ are the 2$\times$2 unit and zero
matrices respectively, and $\omega$ and $\kappa$ are constant multipliers.

Notice that parameter $\kappa$ can be chosen zero since the transformation $%
\Psi\to e^{\mathrm{i}\kappa mt}\Psi$ leaves equation (\ref{equ}) invariant.
Parameter $\omega$ is inessential too since it can be annulled by the
transformation $\beta_\mathtt{m}\to U^\dag \beta_\mathtt{m} U,$ where $%
\mathtt{m}=0,1,2,3,4$ and
\begin{equation*}
U=\left(%
\begin{array}{cc}
\mathtt{I} & -\mathrm{i }\omega \mathtt{I} \\
\mathbf{0} & \mathtt{I}%
\end{array}%
\right).
\end{equation*}

Let us note that if we consider a more general case in which matrices $S_a$
and $\eta_a$ are represented by a direct sum of an arbitrary finite number
of matrices (\ref{02}) and solve the related equations (\ref{cond}), then we
obtain matrices $\beta_\mathtt{m}$ which can be reduced to direct sums of
matrices (\ref{beta}) and zero matrices. In other words, equation (\ref{equ}%
) with matrices (\ref{beta}) is the only non-decoupled system of the first
order equations for spin 1/2 field invariant under the Galilei group.

Equation (\ref{equ}) with matrices (\ref{beta}) and $\omega=\kappa=0$
coincides with the L\'evy-Leblond equation in \cite{levyleblond}.

Let us remark that matrices $\hat\gamma_\mathtt{n}=\eta\beta_\mathtt{n}%
|_{\kappa=\omega=0},\ \mathtt{n}=0,1,2,3,4 $ with
\begin{equation}  \label{geta}
\eta=\left(%
\begin{array}{cc}
\mathbf{0} & \mathtt{I} \\
\mathtt{I} & \mathbf{0}%
\end{array}%
\right)
\end{equation}
satisfy the following relations
\begin{equation}  \label{klifford}
\hat\gamma_\mathtt{n}\hat\gamma_\mathtt{m}+\hat\gamma_\mathtt{m} \hat\gamma_%
\mathtt{n}=2\hat g_{\mathtt{nm}},
\end{equation}
where $\hat g_{\mathtt{nm}}$ is a symmetric tensor whose non-zero components
are
\begin{equation}  \label{g}
\hat g_{04}=\hat g_{40}=-\hat g_{11}=-\hat g_{22}= -\hat g_{33}=1.
\end{equation}

In the Galilei--invariant approach tensor (\ref{g}) plays the same role as
the metric tensor (\ref{gr}) for the Minkovski space in the relativistic theory.

The matrices $\hat\gamma_\mathtt{m}$ will be used many times later on.
Therefore, for convenience, we present them explicitly, namely:
\begin{equation}  \label{gamma}
\begin{array}{c}
\hat\gamma_0=\left(%
\begin{array}{cc}
\mathbf{0} & \mathbf{0} \\
\mathtt{I} & \mathbf{0}%
\end{array}%
\right),\ \hat\gamma_a=\left(%
\begin{array}{cc}
\mathbf{0} & -\sigma_a \\
\sigma_a & \mathbf{0}%
\end{array}%
\right),\ \hat\gamma_4=\left(%
\begin{array}{cc}
\mathbf{0} & 2\mathtt{I} \\
\mathbf{0} & \mathbf{0}%
\end{array}%
\right), \
a=1,2,3.%
\end{array}%
\end{equation}

\subsection{\protect\bigskip Scalar and vector fields and the
corresponding wave equations}


\subsubsection{\label{S23}The indecomposable representations for scalar and
vector fields}

A complete description of indecomposable representations of the algebra $%
hg(1,3)$ in the spaces of vector and scalar fields is given in \cite{Marc}.
The corresponding matrices $S_{a}$ and $\eta _{a}$ have the following forms:
\begin{equation}
\begin{array}{l}
S_{a}=\left(
\begin{array}{cc}
\mathtt{I}_{n\times n}\otimes s_{a} & \cdot \\
\cdot & \mathbf{0_{m\times m}}%
\end{array}%
\right) ,
\ \
\eta _{a}=\left(
\begin{array}{cc}
A_{n\times n}\otimes s_{a} & B_{n\times m}\otimes k_{a}^{\dag } \\
C_{m\times n}\otimes k_{a} & \mathbf{0_{m\times m}}%
\end{array}%
\right) ,%
\end{array}
\label{s}
\end{equation}%
where $\mathtt{I}_{n\times n}$ and $\mathbf{0_{m\times m}}$ are unit and
zero matrices of dimension ${n\times n}$ and ${m\times m,}$ respectively, $%
A_{n\times n}$, $B_{n\times m}$ and $C_{m\times n}$ are matrices of
indicated dimensions whose forms will be specified later on, $s_{a}$ are
matrices of spin one with elements $(s_{a})_{bc}=i\varepsilon _{abc}$ and $%
k_{a}$ are $1\times 3$ matrices of the form
\begin{equation}
k_{1}=\left( \mathrm{i},0,0\right) ,\qquad k_{2}=\left( 0,\mathrm{i}%
,0\right) ,\qquad k_{3}=\left( 0,0,\mathrm{i}\right) .  \label{k}
\end{equation}

Matrices (\ref{s}) fulfill relations (\ref{e33}), iff
matrices $A_{n\times n}, B_{n\times m}$ and $C_{m\times n}$ satisfy the
following relations (we have omitted the related subindices):
\begin{equation}  \label{A1}
AB=0,\ \ CA=0,\ \ A^2+BC=0.
\end{equation}
This system of matrix equations appears to be completely solvable, i.e. it
is possible to find all non-equivalent indecomposable matrices $A, B$ and $C$
which satisfy relations (\ref{A1}). Any set of such matrices generates a
representation of the algebra $hg(1,3)$ whose basis elements are of the
forms specified in (\ref{s}).

According to \cite{Marc} indecomposable representations $D(n,m,\lambda )$ of
$hg(1,3)$ for scalar and vector fields are labelled by integers $n,k$ and $%
\lambda $. They specify dimensions of submatrices in (\ref{s}) and the rank
of matrix $B,$ respectively. As shown in \cite{Marc}, there exist ten
non-equivalent indecomposable representations $D(n,k,\lambda )$ of $hg(1,3)$
which correspond to matrices $A_{n\times n},\ B_{n\times k}$ and $C_{k\times
n}$ given in the Table 1.

In addition to the scalar representation whose generators are written in (%
\ref{s}) and in Item 1 of Table 1, there exist nine vector representations
corresponding to matrices enumerated in Table 1, items 2-10. The
corresponding basis elements are matrices of dimension $(3n+k)\times (3n+k)$
whose explicit forms are given in (\ref{s}) and in Table 1.

\newpage

\begin{center}
Table 1. Solutions of equations (\ref{A1})
\end{center}

\begin{tabular}{|l|l|l|}
\hline
$\mathtt{No}$ & $(n,k,\lambda)$ & $\mathtt{Matrices }A,\ B, C$ \\ \hline
\ 1. & (0,1,0) & $A,\ B\ \texttt{and} \ C\ \texttt{do not exist since }n=0$ \\
\ 2. & (1,0,0) & $A=0$, \ $B\ \texttt{and} \ C\ \texttt{do not exist since }k=0$ \\
\ 3. & (1,1,0) & $A=0,\ B=0,\ C=1$ \\
\ 4. & (1,1,1) & $A=0,\ B=1,\ C=0$ \\
\ 5. & (1,2,1) & $A=0,\ B=(1\ 0),\ C=\left(%
\begin{array}{c}
0 \\
1%
\end{array}%
\right)$ \\
\ 6. & (2,0,0) & $A=\left(%
\begin{array}{cc}
0 & 0 \\
1 & 0%
\end{array}%
\right)$, \ \ $B\ \texttt{and}\ C\ \texttt{do not exist since }k=0$ \\
\ 7. &\ \  (2,1,0) & $A=\left(%
\begin{array}{cc}
0 & 0 \\
1 & 0%
\end{array}%
\right),\ B=\left(%
\begin{array}{c}
0 \\
0%
\end{array}%
\right), \ C=(1\ 0)$ \\
\ 8. & (2,1,1) & $A=\left(%
\begin{array}{cc}
0 & 0 \\
1 & 0%
\end{array}%
\right),\ B=\left(%
\begin{array}{c}
1 \\
0%
\end{array}%
\right), \ C=(0\ 0)$ \\
\ 9. & (2,2,1) & $A=\left(%
\begin{array}{cc}
0 & 0 \\
1 & 0%
\end{array}%
\right), B=\left(%
\begin{array}{cc}
0 & 0 \\
1 & 0%
\end{array}%
\right), C=\left(%
\begin{array}{cc}
0 & 0 \\
1 & 0%
\end{array}%
\right)$ \\
10. & (3,1,1) & $A=\left(%
\begin{array}{ccc}
0 & 0 & 0 \\
1 & 0 & 0 \\
0 & 1 & 0%
\end{array}%
\right),\ \ B=\left(%
\begin{array}{r}
0 \\
0 \\
-1%
\end{array}%
\right)$, \ $C=(1\ 0\ 0)$ \\ \hline
\end{tabular}

\vspace{2mm}

The finite Galilei transformations of vector fields (which can be obtained
by integrating the Lie equations for generators (\ref{s})) and examples of
such fields can be found in paper \cite{Marc}.

\subsubsection{\label{S24}General wave equations for vector and scalar fields}

Let us consider equation (\ref{equ}) and describe all admissible matrices $%
\beta _{4}$ compatible with the invariance conditions (\ref{cond}). We shall
restrict ourselves to matrices $\eta _{a},S_{a}$ belonging to the
representations described in Subsection \ref{S23} (see Table 1) or to direct sums
of these representations. Then the general form of matrices $S_{a}$ and $%
\eta _{a}$ is again given by equations (\ref{s}) where, however, matrices $%
A,B$ and $C$ can be reducible:
\begin{equation}
\begin{array}{c}
A=\left(
\begin{array}{ccccc}
A_{1} &  &  &  &  \\
& A_{2} &  &  &  \\
&  & \cdot &  &  \\
&  &  & \cdot &  \\
&  &  &  & \cdot%
\end{array}%
\right) ,\ B=\left(
\begin{array}{ccccc}
B_{1} &  &  &  &  \\
& B_{2} &  &  &  \\
&  & \cdot &  &  \\
&  &  & \cdot &  \\
&  &  &  & \cdot%
\end{array}%
\right) , \
C=\left(
\begin{array}{ccccc}
C_{1} &  &  &  &  \\
& C_{2} &  &  &  \\
&  & \cdot &  &  \\
&  &  & \cdot &  \\
&  &  &  & \cdot%
\end{array}%
\right)%
\end{array}
\label{n1}
\end{equation}%
and are of dimensions $N\times N,M\times N$ and $N\times M$ respectively
with $N$ and $M$ being arbitrary integers. The unit and zero matrices in the
associated spin operator $\mathbf{S}$ defined by equation (\ref{s}) are $%
(N\times N)-$ and $(M\times M)-$dimensional, respectively.

The sets of matrices $(A_{1},B_{1},C_{1}),\ (A_{2},B_{2},C_{2}),...$ are
supposed to be indecomposable sets presented in Table 1. Any of them is
labelled by a multi-index $q_{i}=(n_{i},k_{i},\lambda _{i}),\ i=1,2,\cdots $.

Matrices $\beta_4$ and $\beta_0$ must commute with $\mathbf{S}$ and
therefore have the following block diagonal form:
\begin{equation}  \label{n2}
\beta_4=\left(%
\begin{array}{cc}
R_{N\times N} & \mathbf{0_{N\times M}} \\
\mathbf{0_{M\times N}} & E_{M\times M}%
\end{array}%
\right),\ \beta_0=\left(%
\begin{array}{cc}
F_{N\times N} & \mathbf{0_{N\times M}} \\
\mathbf{0_{M\times N}} & G_{M\times M}%
\end{array}%
\right).
\end{equation}

Let us denote by $|q,s,\nu >$ a vector belonging to a carrier space of the
representation $D_{q}$ of the algebra $hg(1,3)$, where $q=(n,k,\lambda )$ is
a multi-index which labels a particular indecomposable representation as
indicated in Table 1, $s$ is a spin quantum number which is equal to 0,1 and
index $\nu $ specifies degenerate subspaces with the same fixed $s$. Then
taking into account that matrix $\beta _{4}$ commutes with $S_{a}$ its
elements can be written as
\begin{equation}
\begin{array}{c}
<q,s,\nu |\beta _{4}|q^{\prime },s^{\prime },\nu ^{\prime }>=\delta
_{s1}\delta _{s^{\prime }1}R_{\nu \nu ^{\prime }}(q,q^{\prime }) +\delta _{s0}\delta _{s^{\prime }0}E_{\nu \nu ^{\prime }}(q,q^{\prime }).%
\end{array}
\label{b1}
\end{equation}

In order to find matrices $R(q,q^{\prime })$ and $E(q,q^{\prime })$ (whose
elements are denoted by $R_{\nu \nu ^{\prime }}$ and $E_{\nu \nu ^{\prime }}$
respectively) expression (\ref{b1}) has to be substituted into (\ref{cond})
and matrices $\eta $ in form (\ref{s}) together with relations (\ref{A1})
used. As a result we obtain the following condition
 \beq\label{b2}(A^\dag)^2R+R(A')^2=A^\dag
RA'-C^\dag EC',\eeq
where $A,\ C$ (and $A^{\prime },\ C^{\prime }$) are submatrices used in (\ref%
{A1}), which correspond to representation $D_{q}$ (and $D_{q^{\prime }}$).

Formulae (\ref{b1}) and (\ref{b2}) express all necessary and sufficient
conditions for matrix $\beta _{4}$ imposed by the Galilei invariance
conditions (\ref{cond}). Suppose matrix $\beta _{4}$ in (\ref{b1}) which
satisfies (\ref{b2}) be known, then the remaining matrices $\beta _{a}$ ($%
a=1,2,3$) and $\beta _{0}$ can be found by a direct use of the first and
second relations in (\ref{cond}). By this way we obtain
\begin{equation}
\begin{array}{c}
<q,s,\lambda |\beta _{0}|q^{\prime },s^{\prime },\lambda ^{\prime }>=\delta
_{s1}\delta _{s^{\prime }1}F_{\lambda \lambda ^{\prime }}(q,q^{\prime })
+\delta _{s0}\delta _{s^{\prime }0}G_{\lambda \lambda ^{\prime
}}(q,q^{\prime }), \\
<q,s,\lambda |\beta _{a}|q^{\prime },s^{\prime },\lambda ^{\prime
}>=i(\delta _{s1}\delta _{s^{\prime }1}H_{\lambda \lambda ^{\prime
}}(q,q^{\prime })s_{a}
+\delta _{s1}\delta _{s^{\prime }0}M_{\lambda \lambda ^{\prime
}}(q,q^{\prime })k_{a}^{\dag }\\-\delta _{s0}\delta _{s^{\prime
}1}M_{_{\lambda \lambda ^{\prime }}}^{\dag }(q,q^{\prime })k_{a}),%
\end{array}
\label{b3}
\end{equation}%
where $s_{a}$ are matrices of spin one, $k_{a}$ are matrices (\ref{k}) and $%
F(q,q^{\prime }),\ G(q,q^{\prime }),\ H(q,q^{\prime })$ and $\ M(q,q^{\prime
})$ are matrices defined by the following relations
\beq\label{b4}\bea{c}H=A^\dag
R-RA',\ \ M=C^\dag E-RB',\ \  F=C^\dag EC' +A^\dag RA',\\ G=2B^\dag
RB'-B^\dag C^\dag E-EC'B'. \eea\eeq

Thus, in order to derive a Galilei-invariant equation (\ref{equ}) for vector
fields, it is sufficient to choose a realization of the algebra $hg(1,3)$
from Table 1 or a direct sum of such realizations and find the associated
matrix $\beta _{4}$ (\ref{b1}) whose block matrices $R$ and $E$ satisfy
relations (\ref{b2}). Then the corresponding matrices $\beta _{0}$
and $%
\beta _{a}$ are determined via relations (\ref{b3}) and (\ref{b4}).

All non-trivial solutions of matrices $R$ and $E$ are specified in the
Appendix. Thus \textit{formulae (\ref{b1})- (\ref{b4}) and
Tables 2-4 in the Appendix determined all possible matrices $\beta _{4}$
and $\beta _{\mu }$ which define the Galilei-invariant equations (\ref{equ}) for
fields of spin 1.}

Notice that all these equations admit a Lagrangian formulation with
Lagrangians in the following standard form
\begin{equation}
L=\frac{1}{2}\Psi ^{\dag }\left( \beta _{\mu }p^{\mu }+\beta _{4}m\right)
\Psi +h.c..  \label{lagra}
\end{equation}

\subsubsection{\label{S25}Consistency conditions}

In the previous section we have found all matrices $\beta _{\texttt{m}}$ for
which equation (\ref{equ}) is invariant with respect to vector and scalar
representations of the homogeneous Galilei group. However, the Galilei
invariance itself guarantees neither the consistency
nor the right number of independent components. Moreover spin content of
each of the obtained equations as well as their possibilities to describe
fundamental quantum-mechanical systems have not yet been discussed .

In this subsection we present further constraints on matrices $\beta _{%
\texttt{m}}$ which should be imposed in order to obtain a consistent
equation for a system with a fixed spin.

In this context, a non-relativistic quantum system is said to be \textit{%
fundamental} if the space of its states forms a carrier space of an
irreducible representation of the Galilei group $G(1,3)$. We shall call such
systems as "non-relativistic particles" or simply "particles".

A non-relativistic quantum system is said to be \textit{composed} provided
the space of its states forms a carrier space of some reducible
representation of $G(1,3)$.

First we consider equations (\ref{cov}) for fundamental systems. Notice that
for the group $G(1,3)$ there exist the following three invariant operators:%
\begin{equation}
C_{1}=M,\ C_{2}=2MP_{0}-\mathbf{P}^{2},\ \texttt{and }C_{3}=(M\mathbf{J}-%
\mathbf{P}\times \mathbf{G})^{2}.  \label{cas1}
\end{equation}%
Here, $M$ and $P_{0}$ are scalars and $\mathbf{P,J}$ and $\mathbf{G}$ are
three-vectors whose components are specified by equations (\ref{cov}).
Eigenvalues of the operators $C_{1},C_{2}$ and $C_{3}$ are associated with
mass, internal energy and with square of mass operator multiplied by
eigenvalues of total spin, respectively.

Galilei-invariant equation (\ref{equ}) is said to be \textit{consistent} and
describes a particle with mass $m$, internal energy $\varepsilon $ and spin $%
s$ if it has non-trivial solutions $\Psi $ which form a (non-degenerate)
carrier space of some representation of the Galilei group on which the
following conditions are true:%
\begin{equation}
C_{1}\Psi =m\Psi ,\ \ C_{2}\Psi =\varepsilon \Psi \ \ \texttt{and}\ \ \  C_{3}\Psi
=m^{2}s(s+1)\Psi .  \label{cas2}
\end{equation}%
Moreover, for fundamental particle the spin value $s$ is fixed which
corresponds to an irreducible representation, while for composed system the
corresponding spin operator has two values: $s=0$ and $s=1$ since the
associated representation is reducible.

In accordance with the above definition of consistency, the number of
independent components of function $\Psi ,$ satisfying a consistent equation
(\ref{equ}), must be equal to the number of spin degrees of freedom, namely
to $2s+1$ for a fundamental particle and to 4 for a composed one.

Now we shall show that relations (\ref{cas2}) generate extra conditions for $%
\beta$--matrices so that equation (\ref{equ}) with such $\beta_\texttt{m}$
guarantees the validity of equations (\ref{cas2}).

Function $\Psi $ satisfying equation (\ref{equ}) cotransforms according to a
particular representation of the Galilei group whose infinitesimal
generators are specified in (\ref{cov}) Using (\ref{cov})
we find the following forms of invariant operators (\ref{cas2}):%
\beq\label{cas3}\bea{l}C_1=Im,\ \ C_2=2mp_0-{\bf p}^2,\\
C_3=m^2{\bf S}^2+m({\bf S}\times
{\mbox{\boldmath$\eta$\unboldmath}}) \cdot{\bf
p}-m({\mbox{\boldmath$\eta$\unboldmath}}\times{\bf S} )\cdot{\bf
p}+{\bf p}^2{\mbox{\boldmath$\eta$\unboldmath}}^2- ({\bf
p}\cdot{\mbox{\boldmath$\eta$\unboldmath}})^2.\eea\eeq

We see that $C_3$ is a rather complicated second-order differential operator
with matrix coefficients. In order to diagonalize this operator, we apply
the similarity transformation
\begin{equation}  \label{cas4}
\Psi\to \Psi^{\prime }=W\Psi,\ C_a\to C_a^{\prime }=WC_aW^{-1},\ \ a=1,2,3,
\end{equation}
with
\begin{equation}  \label{W0}
W=\exp\left(\frac{i}{m}{\mbox{\boldmath$\eta$\unboldmath}} \cdot\mathbf{p}%
\right).
\end{equation}

Let us remark that since $({\mbox{\boldmath$\eta$\unboldmath}}\cdot \mathbf{p%
})^{3}=0$ for representations $D(3,1,1)$ and $D(1,2,1)$ and $({%
\mbox{\boldmath$\eta$\unboldmath}}\cdot \mathbf{p})^{2}=0$ for the remaining
representations described in Subsection \ref{S23}, $W$ is the
second- or the first-order differential operator in $\mathbf{x}$.

Using conditions (\ref{e33}) we find that
\begin{equation}
C_{1}^{\prime }=C_{1},\ C_{2}^{\prime }=C_{2}\ \texttt{ and}\ \ C_{3}^{\prime 2}=
\mathbf{S}^{2}.  \label{cas5}
\end{equation}

Function $\Psi ^{\prime }$ has to satisfy conditions (\ref{cas2}) with the
transformed invariant operators (\ref{cas5}), from which follow that
\begin{equation}
(2mp_{0}-\mathbf{p}^{2})\Psi ^{\prime }=\varepsilon \Psi ^{\prime }
\label{cas6}
\end{equation}%
and
\begin{equation}
\mathbf{S}^{2}\Psi ^{\prime }=s(s+1)\Psi ^{\prime }.  \label{cas61}
\end{equation}

In order to see when conditions (\ref{cas6}) and (\ref{cas61}) are true we
transform equation (\ref{equ}) by means of $W$. We obtain
\begin{equation}
\left( \beta _{0}C_{2}+\beta _{4}2m^{2}\right) \Psi ^{\prime }=0
\label{cas7}
\end{equation}%
since, in accordance with (\ref{cond}), we have:%
\begin{equation*}
2m(W^{-1})^{\dag }(\beta ^{\mu }p_{\mu }+\beta _{4}m)W^{-1}=\beta
_{0}(2mp_{0}-\mathbf{p}^{2})+\beta _{4}2m^{2}.
\end{equation*}

Equation (\ref{cas7}) in order to be compatible with (\ref{cas6}) implies
that the matrix $\beta _{0}\varepsilon +\beta _{4}2m^{2}$ (where $%
\varepsilon $ is an eigenvalue of the Casimir operator $C_{2}$) should be
non-regular for some particular value of $\varepsilon $. Moreover, solutions
of equation (\ref{cas7}) must also satisfy condition (\ref{cas61}) and form
a carrier space of irreducible representation $D(s)$ of the rotation group;
consequently, equation (\ref{cas7}) must have $(2s+1)$ independent
solutions. Let us find now restrictions on matrices $\beta _{0}$ and $\beta
_{4}$ generated by these solutions.

As shown in Section \ref{S23}, both matrices $\beta _{0}$ and $\beta _{4}$
have the block diagonal form given by equation (\ref{n2}). Thus equation (%
\ref{cas7}) is decoupled to two subsystems
\begin{equation}
\left( RC_{2}+2m^{2}F\right) \varphi _{1}=0,  \label{cas8}
\end{equation}%
and
\begin{equation}
\left( EC_{2}+2m^{2}G\right) \varphi _{2}=0,  \label{cas9}
\end{equation}%
where the functions $\varphi _{1}$ and $\varphi _{2}$ denote eigenfunction
columns with $3n$ and $k$ components respectively so that $\Psi ^{\prime }=%
\texttt{column}\left( \varphi _{1},\ \varphi _{2}\right) $ .

Equations (\ref{cas6}) and (\ref{cas61}) are consequences of (\ref{cas8})
and (\ref{cas9}) provided determinants of matrices $R,F,E$ and $G$ satisfy
the following conditions:
\begin{equation}
\begin{array}{l}
\texttt{det}\left( RC_{2}+2m^{2}F\right) =\nu C_{2}+\mu , \ \
\texttt{det}\left( EC_{2}+2m^{2}G\right) =\nu ^{\prime }C_{2}+\mu ^{\prime },%
\end{array}
\label{cas10}
\end{equation}%
where $\mu ,\nu $ and $\mu ^{\prime },\nu ^{\prime }$ are numbers satisfying
one of the following relations:
\begin{gather}
\nu \neq 0,\ \mu =-\varepsilon \nu ,\ \nu ^{\prime }=0,\ \mu ^{\prime }\neq
0\ \ \ \ \ \ \texttt{for  }s=1,  \label{MNew} \\
\nu =0,\ \mu \neq 0,\ \nu ^{\prime }\neq 0,\ \mu ^{\prime }=-\varepsilon \nu
^{\prime }\ \ \ \ \ \ \texttt{for  }s=0.
\end{gather}

Thus to find Galilei-invariant equations (\ref{equ}) for particle with a
fixed spin we can start with equations described in the previous section and
impose conditions (\ref{cas10}) on block components of matrices $\beta_0$
and $\beta_4$.

In analogous way we find the consistency conditions for equations (\ref{equ}%
) describing composite systems. In this case relations (\ref{cas6}) and (\ref%
{cas61}) should be valid too. However, $s$ can take two values $s=1$ and $%
s=0 $. We suppose that these spin states are non-degenerate, so that
function $\Psi ^{\prime }$ should have four independent components -- three
of them corresponding to spin one and one to spin zero state. Then
conditions for determinants of matrices $FC_{2}+2m^{2}R$ and $GC_{2}+2m^{2}E$
again have form (\ref{cas10}) where, however,
\begin{equation}
\nu \nu ^{\prime }\neq 0,\ \mu =-\varepsilon \nu ,\ \mu ^{\prime
}=-\varepsilon \nu ^{\prime }.  \label{MMnew}
\end{equation}

In the next sections we consider examples of various Galilean equations (\ref%
{equ}) in more detail.

\section{\label{S3}Special classes of the Galilean wave equations for free
fields}

\subsection{\label{S31}Equations invariant with respect to the
indecomposable representations}

In Sections \ref{S24} and \ref{S25} we have given a general description of all possible
Galilei wave equations (\ref{equ}) for vector and scalar fields. Now we
shall present an analysis of these equations in details. In this and next
sections we shall consider equations for systems with both one fixed spin
state and two spin values states.

First we restrict ourselves to the indecomposable representations of the
algebra $hg(1,3)$ specified in equation (\ref{s}) and Table 1, and find the
associated matrices $\beta _{4},\ \beta _{0}$ and $\beta _{a}$ which appear
in the Galilei-invariant equations (\ref{equ}). Taking into account that $%
A^{\prime }=A,\ B^{\prime }=B$ and $C^{\prime }=C$ in (\ref{b2}), where $A,\
B$ and $C$ are matrices given in Table 1, and using the results present in
Tables 2-4 in the Appendix, we easily find the associated block matrices
$R$, $E$ and consequently all block matrices (\ref{b4}) and matrices (\ref%
{b1}) and (\ref{b3}). In order to simplify matrices $\beta _{\texttt{m}}\ (%
\texttt{m}=0,1,\cdots ,4)$ we use\ the transformations
\begin{equation}
\beta _{\texttt{m}}\rightarrow V^{\dag }\beta _{\texttt{m}}V,  \label{eqtr}
\end{equation}%
where $V$ are invertible matrices commuting with the Galilei boost
generators $\eta _{a}$.

It turns out that non-trivial solutions for $\beta _{4}$ (and consequently
also for $\beta _{0}$ and $\beta _{a}$) exist only for representations $%
D(1,1,0)$, $D(2,1,0)$, $D(2,2,1)$ and $D(3,1,1)$. They have the form:
\begin{gather}\label{M1}\bea{l}
\texttt{For representation } D(1,1,0):\\\beta_4=\left(\bea{cc}\texttt{I}_{3\times3}&\bz_{3\times1}\\
\bz_{1\times3}&0\eea\right),\
\beta_0=\left(\bea{cc}\bz_{3\times3}&\bz_{3\times1}\\
\bz_{1\times3}&2\eea\right),\
\beta_a=\ri\left(\bea{cc}\bz_{3\times3}&-k_a^\dag\\
k_a&0\eea\right);\eea\\\label{M2} \bea{l}
\texttt{For representation }D(2,1,0):\\
\beta_4=\left(\bea{ccc}\bz_{3\times3}&\bz_{3\times3}&\bz_{3\times1}\\
\bz_{3\times3}&\texttt{I}_{3\times3}&\bz_{3\times1}\\\bz_{1\times3}&
\bz_{1\times3}&1 \eea\right),\ \beta_0=\left(\bea{ccc}2
\texttt{I}_{3\times3}
&\bz_{3\times3}&\bz_{3\times1}\\
\bz_{3\times3}&\bz_{3\times3}&\bz_{3\times1}\\\bz_{1\times3}&
\bz_{1\times3}&0
\eea\right),\ \ \ \ \ \  \\\\
\beta_a=\ri\left(\bea{ccc}\bz_{3\times3}
&s_a&k_a^\dag\\
-s_a&\bz_{3\times3}&\bz_{3\times1}\\-k_a& \bz_{1\times3}&0
\eea\right);\eea\end{gather}
\[ \bea{l}
\texttt{For representation }D(2,2,1):\\
\beta_4=\left(\bea{cccc}\bz_{3\times3}&\bz_{3\times3}
&\bz_{3\times1}&\bz_{3\times1}\\
\bz_{3\times3}&\texttt{I}_{3\times3}&\bz_{3\times1}&\bz_{3\times1}\\
\bz_{1\times3}&
\bz_{1\times3}&0&0\\
\bz_{1\times3}& \bz_{1\times3}&0&1 \eea\right),\
\beta_0=\left(\bea{cccc}2 \texttt{I}_{3\times3}&\bz_{3\times3}
&\bz_{3\times1}&\bz_{3\times1}\\
\bz_{3\times3}&\bz_{3\times3}&\bz_{3\times1}&\bz_{3\times1}\\
\bz_{1\times3}&
\bz_{1\times3}&2&0\\
\bz_{1\times3}& \bz_{1\times3}&0&0\eea\right),\eea\]\beq\label{M3}
\beta_a=\ri\left(\bea{cccc}\bz_{3\times3}&s_a
&\bz_{3\times1}&k_a^\dag\\
-s_a&\bz_{3\times3}&-k_a^\dag&\bz_{3\times1}\\
\bz_{1\times3}&
k_a&0&0\\
-k_a& \bz_{1\times3}&0&0 \eea\right);\eeq
\[ \bea{l}
\texttt{And finally for representation }D(3,1,1):\\
\beta_4=\left(\bea{cccc}\bz_{3\times3}&\bz_{3\times3}
&\nu \texttt{I}_{3\times3}&\bz_{3\times1}\\
\bz_{3\times3}&\nu \texttt{I}_{3\times3}&\texttt{I}_{3\times3}&\bz_{3\times1}\\
\nu \texttt{I}_{3\times3}&
\texttt{I}_{3\times3}&\bz_{3\times3}&\bz_{3\times1}\\
\bz_{1\times3}& \bz_{1\times3}&\bz_{1\times3}&-\nu \eea\right),\
\beta_0=\left(\bea{cccc}\bz_{3\times3}&\texttt{I}_{3\times3}
&\bz_{3\times3}&\bz_{3\times1}\\
\texttt{I}_{3\times3}&\bz_{3\times3}&\bz_{3\times3}&\bz_{3\times1}\\
\bz_{3\times3}&
\bz_{3\times3}&\bz_{3\times3}&\bz_{3\times1}\\
\bz_{1\times3}& \bz_{1\times3}&\bz_{1\times3}&0\eea\right),\eea\]
\beq\label{M4}\bea{l}
\beta_a=\ri\left(\bea{cccc}\bz_{3\times3}&\bz_{3\times3}
&s_a&\bz_{3\times1}\\
\bz_{3\times3}&\bz_{3\times3}&\bz_{3\times3}&-k_a^\dag\\
-s_a&
\bz_{3\times3}&\bz_{3\times3}&\bz_{3\times1}\\
\bz_{1\times3}&k_a &\bz_{1\times3}&0 \eea\right).\eea\eeq
Here, $\texttt{I}_{k\times k}$ and $\mathbf{0_{k\times r}}$ are unit and
zero matrices of dimensions ${k\times k}$ and ${k\times r, }${respectively }%
and $\nu $ is an arbitrary non-vanishing parameter.

Thus there are four equations (\ref{equ}) for spinor and vector fields which
are invariant with respect to the above mentioned indecomposable
representations of the homogeneous Galilei group. Their associated matrices $%
\beta_\mu$ and $\beta_4$ are given by formulae (\ref{M1})--(\ref{M4}).

The equations (\ref{equ}) with $\beta $-matrices specified in (\ref{M1}), (%
\ref{M2}), and (\ref{M4}) are equivalent to those discussed in papers \cite{Hagen} and \cite{nikitin},
\cite{fuschich} respectively. However, equation
(\ref{equ}) with $\beta $-matrices of the form (\ref{M3}) is to the best of
our knowledge new. Notice that the related submatrices $F,R$ and $G,E$
satisfy relations (\ref{cas10}) and (\ref{MMnew}), so the associate equation
describes a Galilean quantum-mechanical system whose spin can take two
values: $s=1$ and $s=0$.

Considering matrices (\ref{M4}) we conclude that the corresponding matrices
\begin{equation}
\tilde{\beta}_{\mu }=\eta \beta _{\mu },\ \mu =0,1,2,3,\ \texttt{\ and  }%
\tilde{\beta}_{4}=\eta \beta _{4}-\nu \texttt{I}_{10\times 10},  \label{Nied}
\end{equation}%
where $\eta $ is an invertible matrix
\begin{equation}
\eta =\left(
\begin{array}{cccc}
\mathbf{0_{3\times 3}} & \mathbf{0_{3\times 3}} & \texttt{I}_{3\times 3} &
\mathbf{0_{3\times 1}} \\
\mathbf{0_{3\times 3}} & \texttt{I}_{3\times 3} & \mathbf{0_{3\times 3}} &
\mathbf{0_{3\times 1}} \\
\texttt{I}_{3\times 3} & \mathbf{0_{3\times 3}} & \mathbf{0_{3\times 3}} &
\mathbf{0_{3\times 1}} \\
\mathbf{0_{1\times 3}} & \mathbf{0_{1\times 3}} & \mathbf{0_{1\times 3}} & -1%
\end{array}%
\right)  \label{eta}
\end{equation}%
satisfying the following relations:
\begin{equation}
\beta _{\mu }\beta _{\nu }\beta _{\sigma }+\beta _{\sigma }\beta _{\nu
}\beta _{\mu }=2(g_{\mu \nu }\beta _{\sigma }+g_{\sigma \nu }\beta _{\mu }),
\label{KD2}
\end{equation}%
where $g_{\mu \nu }$ is a "galilean metric tensor" given in (\ref{g}).

Up to definition of metric tensor $g_{\mu\nu}$ relations (\ref{KD2})
coincide with the defining relations for the Duffin-Kemmer-Petiau (DKP)
algebra \cite{kemmer}. Following \cite{fer} we say that such relations
define a Galilean DKP algebra.

Equation (\ref{equ}) with matrices $\tilde{\beta}_{\mu },\tilde{\beta}_{4}$
which satisfy the Galilean DKP algebra is called the {\it
Galilean Duffin-Kemmer equation}. This equation was considered for the first
time apparently in \cite{nikitin}.

There exist also a number of wave equations (\ref{equ}) invariant with
respect to \textit{decomposable} representations of $hg(1,3)$. We shall
discuss some of them in the subsections which follow.

\subsection{\label{S32}The Galilean Proca equations}


In addition to equations discussed in the previous section (whose solutions
belong to carrier spaces of indecomposable representations of the
homogeneous Galilei algebra), there exist also a number of wave equations (%
\ref{equ}) invariant with respect to decomposable representations of $%
hg(1,3) $. Complete description of such equations is given in Section \ref%
{S24} and the Appendix A. In this section we shall discuss some of them as
examples.

Let us consider first wave equations whose solutions are vectors from a
carrier space of the direct sum of representations $D(3,1,1)\oplus D(2,1,1)$
of the algebra $hg(1,3)$. The associated matrices $\beta _{4},\ \beta _{0}$
and $\beta _{a}$ can be found using relations (\ref{b1}), (\ref{b3}), (\ref%
{b4}) and Tables 2 and 3 where submatrices $R(q,q),$ $R(q,q^{\prime }),$ $%
R(q^{\prime },q^{\prime })$ and $E(q,q),$ $E(q,q^{\prime }),$ $E(q^{\prime
},q^{\prime })$ are specified for $q=(3,1,1)$ and $q^{\prime }=(1,2,1).$ We
can simplify these matrices by using the equivalence transformations (\ref%
{eqtr}).

In this way we realize that there are two consistent equations for the
representation $D(3,1,1)\oplus D(2,1,1)$: one describing a particle with
spin $s=1$ and the other describing a composed system with spins $s=1$ and $%
s=0$. Here we present a covariant formulation of these equations.

The equation for a particle with spin $s=1$ can be written in the following
form
\begin{equation}
\begin{array}{l}
p^{\texttt{k}}\Psi ^{\texttt{n}}-p^{\texttt{n}}\Psi ^{\texttt{k}}=m\Psi ^{%
\texttt{kn}}, \\
p_{\texttt{k}}\Psi ^{\texttt{nk}}=\lambda \delta ^{\texttt{n4}}m\Psi ^{4}.%
\end{array}
\label{P12}
\end{equation}%
Here, $\delta ^{\texttt{nk}}$ is the Kronecker symbol, $\Psi ^{\texttt{nk}}\
(\texttt{n,k}=0,1,2,3,4)$ is an antisymmetrical tensor and $\Psi ^{\texttt{n}%
}$ is a five-vector which transform in accordance with representations $%
D(3,1,1)$ and $D(1,2,1),$ respectively.

Equation (\ref{P12}) includes an arbitrary real parameter $\lambda\neq0$
whose value cannot be fixed using only conditions of the Galilei invariance.

The equation for a system with two spin states involves again tensor and
five-vector variables which satisfy the following system:
\begin{equation}
\begin{array}{l}
p^{\texttt{k}}\Psi ^{\texttt{n}}-p^{\texttt{n}}\Psi ^{\texttt{k}}=m\Psi ^{%
\texttt{kn}}, \\
p_{\texttt{k}}\Psi ^{\texttt{nk}}=\nu m\Psi ^{\texttt{n}}.%
\end{array}
\label{P14}
\end{equation}%
The Galilei-invariant equations (\ref{P12}), (\ref{P14}) are quite similar
to the relativistic Proca equations (\ref{001}), but with the following
differences:
\begin{itemize}
\item all indices $\texttt{m,n}$ in (\ref{P12}) take values 0,...,4 while in (\ref%
{001}) we have $\mu ,\nu =0,1,2,3$;
\item the relativistic four-gradient $p_\nu$ is replaced by the Galilean
five-vector $p=(p^0,\ p^1,\ p^2,\ p^3,\ p^4)$, where $p^0=\mathrm{i}\frac{%
\partial}{\partial t},\ p^a=\mathrm{i}\frac\partial{\partial x_a},\ \
a=1,2,3,\ \text{and}\ p^4=m;$

\item covariant indices $\texttt{m}, \texttt{n}$ are raised and lowered by
using the Galilean metric tensor (\ref{g}) instead of the relativistic
metric tensor
\begin{equation}  \label{gr}
g_{\mu\nu}=diag(1,-1,-1,-1);
\end{equation}

\item finally, equation (\ref{P14}) describes a system with two spin states,
while both the relativistic Proca equation and the Galilei-invariant
equation (\ref{P12}) describe a particle with fixed spin $s=1$. We shall
call (\ref{P12}) \textit{the Galilean first order Proca equation}.
\end{itemize}

Let us stress here that the considered Galilean Proca equations are not
non-relativistic approximations of the relativistic Proca equation (\ref{001}%
) (see appendix C where a contraction of (\ref{001}) to its
non-relativistic approximation is shown).

Substituting the expression for $\Psi ^{\texttt{nk}}$ from the first into
the second equation in (\ref{P12}), we obtain the \textit{Galilean second
order Proca equation}:
\begin{equation}
\begin{array}{l}
p_{\texttt{n}}p^{\texttt{n}}\Psi ^{\texttt{m}}-p^{\texttt{m}}p_{\texttt{n}%
}\Psi ^{\texttt{n}}+\lambda \delta ^{\texttt{m}\ 0}m^{2}\Psi ^{4}=0%
\end{array}
\label{gproca1}
\end{equation}%
where $\texttt{m,n}=0,1,2,3,4$. Equation (\ref{gproca1}) admits, like (\ref%
{P12}), a Lagrangian formulation and describes a particle with spin 1. The
corresponding Lagrangian has the form:
\begin{equation}
\begin{array}{c}
L=(p_{\texttt{m}}\Psi _{\texttt{n}}-p_{\texttt{n}}\Psi _{\texttt{m}})^{\ast
}(p^{\texttt{m}}\Psi ^{\texttt{n}}-p^{\texttt{n}}\Psi ^{\texttt{m}}) +(p^{\texttt{m}}\Psi _{\texttt{m}})^{\ast }p_{\texttt{n}}\Psi ^{\texttt{n}}\\
-(p_{\texttt{m}}\Psi _{\texttt{n}})^{\ast }p^{\texttt{m}}\Psi ^{\texttt{n}%
}%
-\lambda m^{2}\Psi _{0}^{\ast }\Psi ^{4},%
\end{array}
\label{lagran}
\end{equation}%
where the asterisk $\ast\ $ denotes complex conjugation.

\subsection{\label{S33}The Galilean Rarita-Schwinger equation}

Till now we have used our knowledge of the indecomposable representations of
the algebra $hg(1,3)$ for spinor, scalar and vector fields to construct wave
equations for fields of spin $\tilde s\leq1$. In this section we derive
Galilean invariant equations for the field transforming as a direct product
of spin 1/2 and spin 1 fields. The relativistic analogue of such system is
the famous Rarita-Schwinger equation.

The relativistic Rarita-Schwinger equation for a particle with spin $s=\frac{%
3}{2}$ is constructed by using a vector-spinor wave function $\Psi _{\alpha
}^{\mu }$, where $\mu =0,1,2,3$ and $\alpha =1,2,3,4$ are vector and spinor
indices respectively. Moreover, $\Psi _{\alpha }^{\mu }$ is supposed to
satisfy the equation
\begin{equation}
\begin{array}{c}
\left( \gamma ^{\nu }p_{\nu }-m\right) \Psi ^{\mu }-\gamma ^{\mu }p_{\nu
}\Psi ^{\nu }-p^{\mu }\gamma _{\nu }\Psi ^{\nu }
+\gamma ^{\mu }\left( \gamma _{\nu }p^{\nu }+m\right) \gamma _{\sigma }\Psi
^{\sigma }=0,%
\end{array}
\label{rs1}
\end{equation}%
where $\gamma ^{\mu }$ are the Dirac matrices acting on the spinor index $%
\alpha $ of $\Psi _{\alpha }^{\mu }$ which we have omitted.

Reducing the left-hand side of equation (\ref{rs1}) by $p_{\mu }$ and $%
\gamma _{\mu }$ we obtain the following expressions:
\begin{equation}
\gamma _{\mu }\Psi ^{\mu }=0,\ \texttt{and }p_{\mu }\Psi ^{\mu }=0,
\label{rs2}
\end{equation}%
which reduce the number of independent components of $\Psi _{\alpha }^{\mu }$
to 8 as required for a wave function of a relativistic particle with spin
3/2.

Using our knowledge of invariants for the Galilean vector fields from \cite%
{Marc} and \cite{NN2} we can easily find a Galilean analogue of equation (%
\ref{rs1}). Like in the case of the Galilean Proca equation we begin with a
five-vector $\Psi ^{\texttt{m}},\ \texttt{m}=0,1,2,3,4$ which has, in
addition, a bi-spinorial index which we do not write explicitly. Thus our
Galilei-invariant equation can be written in the following form:
\begin{equation}
\hat{\gamma}_{\texttt{n}}p^{\texttt{n}}\hat{\Psi}^{\texttt{m}}-\hat{\gamma}^{%
\texttt{m}}p_{\texttt{n}}\Psi ^{\texttt{n}}-p^{\texttt{m}}\hat{\gamma}_{%
\texttt{n}}\Psi ^{\texttt{n}}+\hat{\gamma}^{\texttt{m}}\hat{\gamma}_{\texttt{%
n}}p^{\texttt{n}}\hat{\gamma}_{\texttt{r}}\hat{\Psi}^{\texttt{r}}+\lambda
\delta ^{\texttt{m}0}m\hat{\Psi}^{4}=0.  \label{ra}
\end{equation}%
Here, $\hat{\gamma}_{\texttt{n}}$ are the Galilean $\gamma $-matrices (\ref%
{gamma}), $p^{\texttt{m}}$ is a Galilean "five-momentum" defined in the
previous subsection (see the second Item there), $\lambda $ is an arbitrary
non-vanishing parameter and raising and lowering of indices
$\texttt{m and }%
\texttt{n}$ is made by using the Galilean metric tensor (\ref{g}).

Like equation (\ref{gproca1}) the equation (\ref{ra}) admits a Lagrangian
formulation. The corresponding Lagrangian can be written as
\begin{equation}  \label{lag}
\begin{array}{c}
L=\frac12(\bar{\hat\Psi}_\texttt{m} \hat\gamma_\texttt{n} p^\texttt{n}
\hat\Psi^\texttt{m} -\bar{\hat\Psi}_\texttt{m}\hat\gamma^\texttt{m} p_
\texttt{n}\hat\Psi^\texttt{n}-\bar{\hat\Psi}_\texttt{m} p^\texttt{m}%
\hat\gamma_\texttt{n}\hat\Psi^\texttt{n}
+ \bar{\hat\Psi}_\texttt{m}\hat\gamma^\texttt{m}\hat\gamma_\texttt{n} p^%
\texttt{n}\hat\gamma_\alpha \hat\Psi^\alpha+ \lambda m\bar{\hat\Psi}%
_0\hat\Psi^4)+h.c.,%
\end{array}%
\end{equation}
where $\bar {\hat\Psi}_\texttt{m}= \hat\Psi_\texttt{m}^\dag\eta$ and $\eta$
is the hermitizing matrix (\ref{geta}).

We prove in Appendix B that equation (\ref{ra}) indeed describe a particle
with spin 3/2.

\section{\label{S4}Equations for charged particles interacting with an
external gauge field}

\subsection{\label{S41}Minimal interaction with an external field}

We have described Galilei-invariant equations (\ref{equ}) for free particles
with spins 0, 1/2,1 and 3/2. These equations have admitted Lagrangian
formulation (\ref{lagra}), so that to generalize them to the case of
particles interacting with an external field means, as usually, to apply the
minimal interaction principle, i.e., to make the following change in the
Lagrangian
\begin{equation}
p^{\mu }\rightarrow \pi ^{\mu }=p^{\mu }-eA^{\mu },  \label{minimal}
\end{equation}%
where $A^{\mu }$ are components of a vector-potential of the external field,
and $e$ is a particle charge.

Thus we obtain the Lagrangian
\begin{equation}  \label{minimal_equation}
L=\frac12\Psi^\dag(\mathbf{x}, t)\left(\beta_\mu \pi^\mu+\beta_4m\right)
\Psi(\mathbf{x}, t)+h.c.
\end{equation}

It is important to note that change (\ref{minimal}) is compatible with the
Galilei invariance provided the components $(A^{0},A^{1},A^{2},A^{3})$ of
the vector-potential transform as a Galilean four-vector, i.e., as
\begin{equation}
A^{0}\rightarrow A^{0}+\mathbf{v}\cdot \mathbf{A},\text{ }\ \mathbf{A}%
\rightarrow \mathbf{A}.  \label{trans}
\end{equation}%
Such potential corresponds to a "magnetic" limit of the Maxwell
equations, see \cite{ll1967}, \cite{lebellac}. For other possible
Galilei potentials of the electromagnetic field see \cite{NN3}.

The desired equations for a charged particle interacting with an external
field are the Euler equations derived from the Lagrangian (\ref%
{minimal_equation}).

However let us note that it is also possible to introduce interaction with
an external field via other means, e.g., via an anomalous (Pauli) term.

\subsection{\label{S42}The Galilean Bhabha equations with minimal and
anomalous interactions}

Taking Lagrangian (\ref{minimal_equation}) we can derive the following
equation for a charged particle interacting with an external field
\begin{equation}  \label{minimal_equation1}
\left(\beta_\mu \pi^\mu+\beta_4\pi^4\right) \Psi(\mathbf{x}, t)=0.
\end{equation}

Let us consider now equation (\ref{minimal_equation1}) with general matrices
$\beta_\mu$ and $\beta_4$. If we restrict ourselves to a vector-potential of
magnetic type, i.e., to $A=(A^0,\mathbf{A},0)$, then
\begin{equation}  \label{min}
\pi^0=p^0-eA^0,\ \pi^a=p^a-eA^a,\ \pi^4=m.
\end{equation}

Like free particle equations (\ref{equ}), the equation (\ref%
{minimal_equation1}) is Galilei-invariant provided matrices $\beta _{\mu
},\beta _{4}$ satisfy conditions (\ref{cond}). Moreover the vector-potential
of an external field has to transform according to (\ref{trans}).

Following Pauli \cite{39} we generalize our equation (\ref{minimal_equation1}%
) by adding to it an interaction terms linear in an electromagnetic field
strength. Then we get the equation:
\begin{equation}
\left( \beta _{\mu }\pi ^{\mu }+\beta _{4}m+F\right) \Psi =0,  \label{anint}
\end{equation}%
where
\begin{equation*}
F=\frac{e}{m}(\mathbf{A}\cdot \mathbf{H}+\mathbf{G}\cdot \mathbf{E}).
\end{equation*}%
Here, $\mathbf{A}$ and $\mathbf{G}$ are matrices determined by requirement
of the Galilei invariance, i.e., by demanding that $\mathbf{A}\cdot \mathbf{H%
}$ and $\mathbf{G}\cdot \mathbf{E}$ have to be Galilean scalars.

In paper \cite{Marc} we have found the most general form of the Pauli
interaction which can be introduced into the L\'{e}vy-Leblond equation for a
particle of spin 1/2 . Finding the general Pauli interaction for other
Galilean particles is a special problem for any equation previously
considered. Here we restrict ourselves to a systematic analysis of the Pauli
terms which is valid for \textit{any} Galilean Bhabha equation.

First we shall prove the following statement.

\textbf{Lemma}. \textit{Let $S_{a},\eta _{a}$ be matrices which realize a
representation of the algebra $hg(1,3)$, $\Lambda $ be a matrix satisfying
the conditions
\begin{equation}
S_{a}\Lambda =\Lambda S_{a},\ \ \eta _{a}^{\dag }\Lambda =\Lambda \eta _{a}
\label{L}
\end{equation}%
and
\begin{equation*}
E_{a}=\frac{\partial A_{0}}{\partial x^{a}}-\frac{\partial A_{a}}
{\partial t%
},\ \ H_{a}=\varepsilon _{abc}\frac{\partial A^{b}}{\partial x_{c}}
\end{equation*}%
be vectors of the electric and magnetic field strength, respectively. Then
matrices
\begin{equation}
F_{1}=\Lambda (\mathbf{s}\cdot \mathbf{H}-{\mbox{\boldmath$\eta$\unboldmath}}%
\cdot \mathbf{E})\ \ and\ \ F_{2}=\Lambda {\mbox{\boldmath$\eta$\unboldmath}}%
\cdot \mathbf{H}  \label{paul5}
\end{equation}%
are invariant with respect to the Galilei transformations provided the
vector-potential $A$ is transformed in accordance with the Galilean law (\ref%
{trans})}.

\textbf{Proof}. First we note that matrices (\ref{paul5}) are scalars with
respect to rotations. Then, starting with transformation laws (\ref{11}) and
(\ref{trans}) we easily find that under a Galilei boost the vectors $\mathbf{%
E}$ and $\mathbf{H}$ co-transform as
\begin{equation}  \label{cotrans}
\mathbf{E}\to \mathbf{E}-\mathbf{v}\times\mathbf{H},\ \ \ \mathbf{H}\to
\mathbf{H}.
\end{equation}

On the other hand the transformation laws for matrices $\Lambda \mathbf{S}$
and ${\Lambda \mbox{\boldmath$\eta$\unboldmath}}$ can be found using the
exponential mapping of boost generators $\mathbf{G}$ given in equation (\ref%
{cov}):
\begin{equation}
\begin{array}{c}
\mathbf{S}\rightarrow \exp (i\mathbf{G}^{\dag }\cdot \mathbf{v})\Lambda
\mathbf{S}\exp (-i\mathbf{G}\cdot \mathbf{v})\\=\Lambda \exp (i{%
\mbox{\boldmath$\eta\cdot v$
\unboldmath}})\mathbf{S}\exp (-i{\mbox{\boldmath$\eta\cdot v$ \unboldmath}})=%
\mathbf{s}+\mathbf{v}\times {\mbox{\boldmath$\eta$\unboldmath}}, \\
{\mbox{\boldmath$\eta$\unboldmath}}\rightarrow \exp (i\mathbf{G}^{\dag
}\cdot \mathbf{v})\Lambda {\mbox{\boldmath$\eta$\unboldmath}}\exp (i\mathbf{G%
}\cdot \mathbf{v})\\=\Lambda \exp (i{%
\mbox{\boldmath$\eta\cdot v$
\unboldmath}}){{\mbox{\boldmath$\eta$
\unboldmath}}}\exp (-i{\mbox{\boldmath$\eta\cdot v$
\unboldmath}})=\Lambda {\mbox{\boldmath$\eta$ \unboldmath}}.%
\end{array}
\label{trah}
\end{equation}

One easily verifies that transformations (\ref{cotrans}) and (\ref{trah})
leave matrices $F_1$ and $F_2$ invariant. Q.E.D.

In accordance with the Lemma there are many possibilities how to generalize
equations (\ref{minimal_equation1}) to the case with anomalous interaction.
Indeed, for any Galilean Bhabha equation there are matrices $S_a$, $\eta_a$
and $\Lambda$ for which conditions (\ref{L}) are satisfied. For example, we
can choose $\Lambda=\beta_0$. In addition, for many cases there exist a
hermitizing matrix $\eta=\Lambda$ satisfying (\ref{L}), see, e.g., equations
(\ref{geta}) and (\ref{eta}). For particular representations of the algebra $%
hg(1,3)$ there are also other solutions of equations (\ref{L}).

Thus the Pauli term for a Galilean Bhabha equation can be chosen
in the form (\ref{paul5}) or, more generally, as a linear combination of
both, $F_{1}$ and $F_{2}$. As a result we obtain the following equation,
\begin{equation}
\begin{array}{c}
Q\Psi \equiv \left( \beta _{\mu }\pi ^{\mu }+\beta _{4}m+\lambda _{1}\frac{e}{m}\beta _{0}{\mbox{\boldmath$\eta$ \unboldmath}}%
\cdot \mathbf{H}+\lambda _{2}\frac{e}{m}\beta _{0}\left( \mathbf{S}\cdot
\mathbf{H}-{\mbox{\boldmath$\eta$ \unboldmath}}\cdot \mathbf{E}\right)
\right) \Psi =0,%
\end{array}
\label{anint1}
\end{equation}%
where $\lambda _{1}$ and $\lambda _{2}$ are dimensionless coupling constants.

Since equation (\ref{anint1}) is reduced to equation (\ref{minimal_equation1}%
) for $\lambda_1=\lambda_2=0$, equation (\ref{anint1}) describes anomalous
as well as minimal interaction.

In order to receive the physical content of this equation it is convenient
to apply the transformation
\begin{equation}  \label{trans1}
\Psi\to \Psi^{\prime}=W^{ -1}\Psi, \ Q \to Q^{\prime}=W^{\dag }QW,
\end{equation}
where
\begin{equation}  \label{W}
W=\exp\left(- i\frac{\mbox{\boldmath$\eta\cdot\pi$\unboldmath}}{m} \right)
\end{equation}
and ${\mbox{\boldmath$\eta$\unboldmath}}$ is a vector whose components are
the Galilei boost generators (\ref{02}). For the case $e=0$ (or $A_\mu=0$)
the operator $W$ reduces to operator $U$ given in equation (\ref{W0}), which
was used for our analysis of free particle equations.

Using relations (\ref{cond}) and supposing that the nilpotence index $N$ of
matrix ${\mbox{\boldmath$\eta\cdot\pi$\unboldmath}}$ satisfies $N<4$ we
obtain the following equation:
\begin{equation}\label{aprox}\begin{array}{l}Q'\Psi'\equiv\left\{(\beta_0\left(\pi^0-\frac{
{\mbox{\boldmath$\pi$\unboldmath}}^2}{2m}+
\frac{e}{m}{\mbox{\boldmath$\eta$\unboldmath}}\cdot {\bf
F}\right)-\frac{e}{2m}{\mbox{\boldmath$\beta$\unboldmath}}
\times{\mbox{\boldmath$\eta$\unboldmath}}\cdot{\bf H}+\beta_4m
-\frac{e}{6m^2}\widehat Q_{ab}\frac{\partial H_a}{\partial
x_b}\right.\\\left. +\frac em\Lambda\left[\lambda_1
{\mbox{\boldmath$\eta$ \unboldmath}}\cdot{\bf H}+\lambda_2\left(
{\bf S}\cdot{\bf H}-{\mbox{\boldmath$\eta$ \unboldmath}}\cdot{\bf
F}+\frac 1{2m}\tilde Q_{ab}\frac{\partial H_a}{\partial
x_b}\right)\right]\right\}\Psi'=0\end{array}\end{equation}
which is equivalent to (\ref{anint1}). Here $\mathbf{E}%
=-\nabla A^{0}-\frac{\partial \mathbf{A}}{\partial t}$ and $\mathbf{H}%
=\nabla \times \mathbf{A}$ are vectors of the corresponding electric and
magnetic field strength respectively,
\begin{equation}
\begin{array}{l}
\mathbf{F}=\mathbf{E}+\frac{1}{2m}({\mbox{\boldmath$\pi$%
\unboldmath}}\times \mathbf{H}-\mathbf{H}\times {\mbox{\boldmath$\pi$%
\unboldmath}})
\end{array}
\label{quadr}
\end{equation}%
and
\begin{equation*}
\tilde{Q}_{ab}=\frac{1}{2}(\eta _{a}S_{b}+\eta _{b}S_{a}+S_{b}\eta _{a}+\eta
_{a}S_{b}), \ \ \widehat{Q}_{ab}=\eta _{a}^{\dag }\varepsilon _{bcd}\beta
_{c}\eta _{d}+\varepsilon _{bcd}\beta _{c}\eta _{d}\eta _{a}.
\end{equation*}

Equation (\ref{aprox}) includes the Schr\"odinger terms $(\pi^0- \frac{ {%
\mbox{\boldmath$\pi$\unboldmath}}^2}{2m})\Psi^{\prime }$ and additional
terms which are linear in vectors of the external field strengthes and their
derivatives.

Notice that if the nilpotence index $N$ of matrices ${\mbox{$\eta_a$%
\unboldmath}}$ satisfies the relation $N<4,$ which is fulfilled for all
representations of algebra $hg(1,3)$ considered in the present paper, the
transformed equation (\ref{aprox}) is completely equivalent to initial
equation (\ref{anint1}).


\subsection{\label{S43}The Galilean equation for spinor particle interacting
with an external field}

Let us consider equation (\ref{aprox}) for two particular realizations of $%
\beta$--matrices in more detail. First notice, that our conclusions from
equations (\ref{minimal_equation1})--(\ref{aprox}) are true in general and
in particular for the L\'evy-Leblond equation, i.e., when $\beta_\mu,
\beta_4 $ are $4\times4$ matrices determined by relations (\ref{beta}) with $%
\omega=\kappa=0$. Then $\beta_0\eta_a=0,\ {\widehat Q}_{ab}=\tilde Q_{ab}=0, \ {%
\mbox{\boldmath$\beta$\unboldmath}} \times{\mbox{\boldmath$\eta$\unboldmath}}%
=-2\beta_0\mathbf{S}$, and equation (\ref{aprox}) is reduced to the
following form:
\begin{equation}  \label{aprox1}
\begin{array}{l}
\left\{\beta_0\left(\pi^0-\frac{ {\mbox{\boldmath$\pi$\unboldmath}}^2}{2m}+%
\frac{e}{m}\mathbf{S}\cdot\mathbf{H}\right)+\beta_4m +\frac em\Lambda\left[\lambda_1 {\mbox{\boldmath$\eta$
\unboldmath}}\cdot\mathbf{H}+\lambda_2\left( \mathbf{S}\cdot\mathbf{H}-{%
\mbox{\boldmath$\eta$ \unboldmath}}\cdot\mathbf{F}\right)\right]%
\right\}\Psi^{\prime }=0.%
\end{array}%
\end{equation}

For $\lambda _{1}=\lambda _{2}=0$ (i.e., when only the minimal interaction
is present) equation (\ref{aprox1}) is reduced to the following system:
\begin{equation}
\left( \pi _{0}-\frac{\mbox{\boldmath$\pi$\unboldmath}^{2}}{2m}+\frac{e}{2m}{%
\mbox{\boldmath$\sigma$\unboldmath}}\cdot \mathbf{H}\right) \varphi _{1}=0
\label{pauli}
\end{equation}

and%
\begin{equation}
m\varphi _{2}=0,\ \ \texttt{or  }\varphi _{2}=0,  \label{var}
\end{equation}%
where $\varphi _{1}=\beta _{0}\Psi ^{\prime }$ and $\varphi _{2}=(1-\beta
_{0})\Psi ^{\prime }$ are two--component spinors.

Thus introducing the minimal interaction (\ref{minimal}) into the
L\'evy-Leblond equation, we get the Pauli equation for physical components
of the wave function; moreover, the coupling constant (gyromagnetic ratio)
for the Pauli interaction $\frac{e}{2m}\hat{\mathbf{s}}\cdot \mathbf{H}$, $%
\hat{\mathbf{s}}=\frac12{\mbox{\boldmath$\sigma$\unboldmath}}$ has the same
value $g=2$ as in the case of the Dirac equation \cite{levyleblond}.

Considering now the case with an anomalous interaction we conclude that the
general form of matrix $\Lambda $ satisfying relations (\ref{L}) is
\begin{equation}  \label{n7}
\Lambda=\nu\beta_0+\mu\eta,
\end{equation}
where $\eta$ is hermitizing matrix (\ref{geta}), $\nu$ and $\mu$ are
arbitrary parameters. Substituting (\ref{n7}) into (\ref{aprox1}) we obtain
the following generalization of system (\ref{pauli}):
\begin{equation}  \label{paulii}
\begin{array}{l}
\left(\pi_0-\frac{\mbox{\boldmath$\pi$\unboldmath}^2} {2m}+\frac{eg}{2m}{%
\mbox{\boldmath$\sigma$\unboldmath}}\cdot \mathbf{H}-\frac{e\lambda_3}{2m}{%
\mbox{\boldmath$\sigma$\unboldmath}}\cdot \mathbf{F}-\frac{\lambda_3^2e^2}{%
8m^3}\mathbf{H}^2\right) \varphi_1=0, \\
\varphi_2=-\frac{\lambda_3e}{4m} {\mbox{\boldmath$\sigma$\unboldmath}}\cdot
\mathbf{H}\varphi_1,%
\end{array}%
\end{equation}
where $g=2+\mu\lambda_1+\nu\lambda_2$ and $\lambda_3=\mu\lambda_2$ are
arbitrary parameters.

We see that the L\'evy-Leblond equation with minimal and anomalous
interactions reduces to the Galilean Schr\"odinger-Pauli equation (\ref%
{paulii}) which, however, includes two
additional terms linear in strengths of an electric field and linear and
quadratic in strengths of a magnetic field.
We shall discuss them in detail
in Section 4.6.

\subsection{\label{S44}The Galilean Duffin-Kemmer equation for particle
interacting with an external field}

Consider now the Galilean Duffin-Kemmer equation for spin one particles
interacting with an external field. The corresponding $\beta $-matrices in (%
\ref{anint1}) have dimension $10\times 10$ and are given explicitly in (\ref%
{M4}). Thus
\beq
\begin{array}{c}\label{AN1}
\beta _{0}{\mbox{\boldmath$\eta$\unboldmath}}=\left(
\begin{array}{cccc}
\mathbf{s} & \mathbf{0_{3\times 3}} & \mathbf{0_{3\times 3}} & \mathbf{%
0_{3\times 1}} \\
\mathbf{0_{3\times 3}} & \mathbf{0_{3\times 3}} & \mathbf{0_{3\times 3}} &
\mathbf{0_{3\times 1}} \\
\mathbf{0_{3\times 3}} & \mathbf{0_{3\times 3}} & \mathbf{0_{3\times 3}} &
\mathbf{0_{3\times 1}} \\
\mathbf{0_{1\times 3}} & \mathbf{0_{1\times 3}} & \mathbf{0_{1\times 3}} & 0%
\end{array}%
\right), \
\beta _{0}\mathbf{S}=\left(
\begin{array}{cccc}
\mathbf{0_{3\times 3}} & \mathbf{s} & \mathbf{0_{3\times 3}} & \mathbf{%
0_{3\times 1}} \\
\mathbf{s} & \mathbf{0_{3\times 3}} & \mathbf{0_{3\times 3}} & \mathbf{%
0_{3\times 1}} \\
\mathbf{0_{3\times 3}} & \mathbf{0_{3\times 3}} & \mathbf{0_{3\times 3}} &
\mathbf{0_{3\times 1}} \\
\mathbf{0_{1\times 3}} & \mathbf{0_{1\times 3}} & \mathbf{0_{1\times 3}} & 0%
\end{array}%
\right), \\
\\\widehat{Q}_{ab}=\left(
\begin{array}{cccc}
-3Q_{ab} & \mathbf{0_{3\times 3}} & \mathbf{0_{3\times 3}} & \mathbf{%
0_{3\times 1}} \\
\mathbf{0_{3\times 3}} & \mathbf{0_{3\times 3}} & \mathbf{0_{3\times 3}} &
\mathbf{0_{3\times 1}} \\
\mathbf{0_{3\times 3}} & \mathbf{0_{3\times 3}} & \mathbf{0_{3\times 3}} &
\mathbf{0_{3\times 1}} \\
\mathbf{0_{1\times 3}} & \mathbf{0_{1\times 3}} & \mathbf{0_{1\times 3}} & 0%
\end{array}%
\right),\
\tilde{Q}_{ab}=\left(
\begin{array}{cccc}
Q_{ab}^{\prime } & \mathbf{0_{3\times 3}} & \mathbf{0_{3\times 3}} & \mathbf{%
0_{3\times 1}} \\
\mathbf{0_{3\times 3}} & \mathbf{0_{3\times 3}} & \mathbf{0_{3\times 3}} &
\mathbf{0_{3\times 1}} \\
\mathbf{0_{3\times 3}} & \mathbf{0_{3\times 3}} & \mathbf{0_{3\times 3}} &
\mathbf{0_{3\times 1}} \\
\mathbf{0_{1\times 3}} & \mathbf{0_{1\times 3}} & \mathbf{0_{1\times 3}}
& 0%
\end{array}%
\right),\eea\eeq\[\bea{c}%
{\mbox{\boldmath$\beta$\unboldmath}}\times {\mbox{\boldmath$\eta$\unboldmath}%
}\cdot \mathbf{H}=-\left(
\begin{array}{cccc}
\mathbf{0_{3\times 3}} & \mathbf{s}\cdot \mathbf{H} & \mathbf{%
0_{3\times 3}} & 2\mathbf{k}^{\dag }\cdot \mathbf{H} \\
\mathbf{s}\cdot \mathbf{H} & \mathbf{0_{3\times 3}} & \mathbf{0_{3\times 3}}
& \mathbf{0_{3\times 1}} \\
\mathbf{0_{3\times 3}} & \mathbf{0_{3\times 3}} & \mathbf{0_{3\times 3}} &
\mathbf{0_{3\times 1}} \\
2\mathbf{k}\cdot \mathbf{H} & \mathbf{0_{1\times 3}} & \mathbf{0_{1\times 3}}
& 0%
\end{array}%
\right) ,
\end{array}%
\]
 where
\begin{equation}
Q_{ab}=s_{a}s_{b}+s_{b}s_{a}-\frac{4}{3}\delta _{ab}\ \text{\ and \ }%
Q_{ab}^{\prime }=Q_{ab}+\frac{4}{3}\delta _{ab}.  \label{Q_}
\end{equation}%
Let us consider the corresponding equation (\ref{aprox}) and
restrict ourselves to $\Lambda =\beta _{0}$. Representing $\Psi
^{\prime }$ as a column vector $(\psi _{1},\psi _{2},\psi
_{3},\varphi ),$ where $\psi _{1},\psi _{2},\psi _{3}$ are
three-component vector functions and $\varphi $ is a one-component
scalar function and using (\ref{M4}), (\ref{AN1}), (\ref{Q_}) we
reduce (\ref{aprox}) to the following Pauli-type equation for $\psi
_{1}$:
\begin{equation}
i\frac{\partial }{\partial t}\psi _{1}=\widehat{H}\psi _{1},  \label{paul3}
\end{equation}%
where
\begin{equation}
\begin{array}{l}
\label{Ha}\widehat{H}=\frac{\nu ^{2}}{2}m+\frac{{\mbox{\boldmath$\pi$%
\unboldmath}}^{2}}{2m}+eA_{0}-\frac{ge}{2m}\mathbf{s}\cdot \mathbf{H}+\frac{%
qe}{\nu m}\mathbf{s}\cdot \mathbf{E}-\frac{qe}{2\nu m^{2}}%
\mathbf{s}\cdot ({\mbox{\boldmath$\pi$\unboldmath}}\times \mathbf{H}-\mathbf{%
H}\times {\mbox{\boldmath$\pi$\unboldmath}}) \\\\
+\frac{e}{2\nu m^{2}}%
(2-q)Q_{ab}\frac{\partial H_{a}}{\partial x_{b}}+\frac{e^{2}}{2\nu ^{2}m^{3}}%
\left( \mathbf{H}^{2}-(\mathbf{s}\cdot \mathbf{H})^{2}\right)\end{array}\eeq \\
and $g=1+2\lambda _{1}+2\lambda _{2},%
\text{ }\text{ }q=1-\lambda _{2}.$

The remaining components of $\Psi ^{\prime }$ can be expressed in terms of $%
\psi _{1}$:
\begin{equation*}
\begin{array}{c}
\varphi =-\frac{e}{\nu ^{2}m^{2}}\mathbf{k}\cdot \mathbf{H}\psi _{1},\ \text{%
\ }\psi _{2}=-\frac{1}{\nu }\psi _{1}, \ \
\psi _{3}=-\nu \psi _{2}-\frac{1}{m}\left( \pi _{0}-\frac{1}{2m}{%
\mbox{\boldmath$\pi$\unboldmath}}^{2}+\frac{e}{2m}\mathbf{s}\cdot \mathbf{H}%
\right) \psi _{1}.%
\end{array}%
\end{equation*}

Notice, that in comparison with (\ref{paulii}) the equation (\ref{paul3})
has an essentially new feature. Namely, excluding anomalous interaction,
i.e., setting $\lambda _{1}=\lambda _{2}=0$ in (\ref{Ha}) it still includes
the term $-\frac{e}{\nu m}\mathbf{s}\cdot \mathbf{E}$ describing the
coupling of spin with an electric field. We shall show in the next section
that this effectively represents the spin-orbit coupling. The other terms of
Hamiltonian (\ref{Ha}) (which are placed in the second line of equation (\ref%
{Ha})) can be neglected starting with a reasonable assumption about possible
values of the magnetic field strength.

However, equation (\ref{paulii}) for $\lambda_1=\lambda_2=0$ reduces to the
Schr\"odinger-Pauli equation (\ref{pauli}) which has nothing to do with the
spin-orbit coupling.

\subsection{\label{S45}The Galilean Proca equation for interacting particles}

Equations (\ref{P12}) and (\ref{P14}) can be generalized too by introducing
minimal and anomalous interactions with an external field. As a result we
obtain:
\begin{equation}
\begin{array}{l}
\pi ^{\texttt{k}}\Psi ^{\texttt{n}}-\pi ^{\texttt{n}}\Psi ^{\texttt{k}%
}=m\Psi ^{\texttt{kn}}, \\
\pi _{\texttt{k}}\Psi ^{\texttt{nk}}=\nu m\Psi ^{\texttt{n}}+\mathrm{i}\mu
\frac{e}{m}F^{\texttt{nk}}\Psi _{\texttt{k}}+\lambda \delta ^{\texttt{n4}%
}m\Psi ^{4},%
\end{array}
\label{P21}
\end{equation}%
where parameters $\lambda $ and $\nu $ satisfy the conditions $\lambda \nu
=0,\ \nu ^{2}+\lambda ^{2}\neq 0$. Formulae (\ref{P21}) generalize both
equation (\ref{P12}) (which corresponds to $\nu =0,\lambda \neq 0$) and
equation (\ref{P14}) (for which $\nu \neq 0,\lambda =0$).

Multiplying equations (\ref{P21}) by both $\pi _{\texttt{n}}$ and $\pi _{%
\texttt{n}}\pi _{\texttt{k}}$, summing up by $m$ and $k$ and then expressing
$\Psi ^{\texttt{nk}},\Psi ^{0}$ via $\Psi ^{a}$ and $\Psi ^{4}$, we obtain
the following system:
\begin{equation}
\begin{array}{l}
(\pi _{0}-\frac{\mathbf{\pi }^{2}}{2m}+\frac{1+2\mu }{2m}\mathbf{s\cdot H}-%
\frac{\nu }{2}m)\Psi+\frac{e}{2m}(1-\mu )%
\mathbf{k}^{\dag }\cdot \mathbf{F}\Psi ^{4}=0, \\
\\
(\nu \pi _{0}+\frac{(\lambda -\nu ^{2})}{2m})\Psi ^{4}-\frac{e}{2m}(1-\mu )%
\mathbf{k}\cdot \mathbf{F}\Psi =0,%
\end{array}
\label{P22}
\end{equation}%
where we have denoted $\Psi =\texttt{column}(\Psi ^{1},\Psi ^{2},\Psi ^{3})$.

Let $\nu =0$ then solving the second of equations (\ref{P22}) for $\Psi ^{4}$
and substituting the result into the first equation we obtain
\begin{equation}
\begin{array}{c}
\left( \pi _{0}-\frac{\mathbf{\pi }^{2}}{2m}+(1+2\mu )\frac{e}{2m}\mathbf{%
s\cdot H}-\frac{\nu }{2}m+(1-\mu )^{2}\frac{e^{2}}{2m^{3}}(\mathbf{F}%
^{2}-(\mathbf{s}\cdot \mathbf{F})^{2}\right) \Psi =0.%
\end{array}
\label{P23}
\end{equation}

In accordance with (\ref{P23}) the Galilean Proca equation for a particle
with spin 1 interacting with an external field is reduced to the Schr\"{o}dinger-Pauli equation with an extra term $~\frac{e^{2}}{2m^{3}}(\mathbf{F}%
^{2}-(\mathbf{s}\cdot \mathbf{F})^{2})$ which can be treated as a small
correction.

For $\nu \neq 0$ equations (\ref{P22}) describe a composed system with spins
$s=1$ and $s=0$. Notice that there are two privileged values of the
arbitrary parameter $\nu $, namely, $\nu =-1$ and $\nu =1.$

If $\nu =-1$ or $\nu =1$ then equations (\ref{P22}) can be rewritten in the
Schr\"{o}dinger form (\ref{paul3}), where $\psi _{1}=\texttt{column}(\Psi
^{1},\Psi ^{2},\Psi ^{3},\Psi ^{4})$, and
\begin{equation}
\hat{H}=\frac{\mathbf{\pi }^{2}}{2m}+eA_{0}-\frac{ge}{2m}\mathbf{S}\cdot
\mathbf{H}-\frac{3-g}{2m}\mathbf{K}\cdot \mathbf{F}-\frac{m}{2}  \label{P24}
\end{equation}%
or
\begin{equation}
\hat{H}=\frac{\mathbf{\pi }^{2}}{2m}+eA_{0}-\frac{ge}{2m}\mathbf{S}\cdot
\mathbf{H}-\frac{3-g}{2m}\mathbf{\hat{K}}\cdot \mathbf{F}+
\frac{m}{2},
\label{P25}
\end{equation}%
respectively. Here $g=1+2\mu $ and $\mathbf{S}$, $\mathbf{K}$ $\mathbf{\hat{K%
}}$ are matrix vectors whose components are
\begin{equation}
\begin{array}{c}
S_{a}=\left(
\begin{array}{cc}
s_{a} & \mathbf{0_{3\times 1}} \\
\mathbf{0_{1\times 3}} & 0%
\end{array}%
\right) ,\ K_{a}=\left(
\begin{array}{cc}
\mathbf{0_{3\times 3}} & k_{a}^{\dag } \\
k_{a} & 0%
\end{array}%
\right) , \ \
{\hat{K}_{a}}=\left(
\begin{array}{cc}
\mathbf{0_{3\times 3}} & k_{a}^{\dag } \\
-k_{a} & 0%
\end{array}%
\right)%
\end{array}
\label{P26}
\end{equation}%
and $k_{a}$ are matrices (\ref{k}). Matrices $\{S_{a},K_{a}\}$ and $\{S_{a},%
\hat{K}_{a}\}$ form bases of the algebra $so(4)$ and $so(1,3),$ respectively.

Notice that Hamiltonian (\ref{P24}) is formally Hermitian w.r.t. the
standard scalar product for the direct sum of four square integrable
functions while (\ref{P25}) is Hermitian w.r.t. the indefinite metric $(\psi
_{1},\psi _{2})=\int \psi _{1}^{\dag }M\psi _{2}d^{3}x,$ where $M$ is either
a matrix whose elements are given in equation (\ref{gr}) or a parity
operator. For other non-vanishing values of the parameter $\nu $ in (\ref%
{P21}), i.e., for $\nu \neq 0,\pm 1$, the corresponding Hamiltonian $\hat{H}$
is non-Hermitian.

\subsection{\label{S46}The Galilei invariance and spin-orbit coupling}

Consider now the first of equations (\ref{paulii}) for particular values of
arbitrary parameters, namely for $\tilde \lambda_1+\tilde \lambda_2=-1$:
\begin{equation}  \label{paulik}
\hat L\varphi_1\equiv\left(\pi_0-\frac{\mbox{\boldmath$\pi$\unboldmath}^2} {%
2m}-\frac{e\lambda_3}{2m}{\mbox{\boldmath$\sigma$\unboldmath}}\cdot \mathbf{F%
}-\frac{\lambda_3^2e^2}{8m^3}\mathbf{H}^2\right) \varphi_1=0.
\end{equation}

First, let us remind that this equation is a direct consequence of the
Galilei-invariant L\'{e}vy-Leblond equation with an anomalous interaction,
i.e., of equation (\ref{anint1}) where $\beta _{\texttt{n}}$ are matrices (%
\ref{beta}) with $\kappa =\omega =0$. Secondly, equation (\ref{paulik}) by
itself is transparently Galilei-invariant since the operator $\hat{L}$ in (%
\ref{paulik}) is a Galilean scalar provided the value of an arbitrary parameter
$\lambda _{3}$ is finite. We shall assume $\lambda _{3}$ to be small.

In order to find out the physical content of equation (\ref{paulik}) we
transform it to a more transparent form using the operator $U=\exp (-\frac{%
\mathrm{i}\lambda _{3}}{2m}{\mbox{\boldmath$\sigma\cdot\pi$\unboldmath}})$.
Applying this operator to $\varphi _{1}$ and transforming $\hat{L}%
\rightarrow \hat{L}^{\prime }=U\hat{L}U^{-1}$ we obtain the equation
\begin{equation}
\begin{array}{c}
L^{\prime }\varphi _{1}^{\prime }=\left( \pi _{0}-\frac{{%
\mbox{\boldmath$\pi$
\unboldmath}}^{2}}{2m}-eA_{0}-\frac{e\lambda _{3}^{2}}{8m^{2}}\left( {\mbox{\boldmath$\sigma$%
\unboldmath}}\cdot ({\mbox{\boldmath$\pi$\unboldmath}}\times \mathbf{E}-%
\mathbf{E}\times {\mbox{\boldmath$\pi$\unboldmath}})-\texttt{div}\mathbf{E}%
\right) +\cdots \right) \varphi _{1}^{\prime }=0,%
\end{array}
\label{orb}
\end{equation}%
where the dots denote small terms of the order $o(\lambda _{3}^{3})$ and $%
o(e^{2})$.

All terms in big round brackets have an exact physical meaning. They include
first the Schr\"{o}dinger terms $\pi _{0}-\frac{{\mbox{\boldmath$\pi$%
\unboldmath}}^{2}}{2m}-eA_{0},$ then the term $\sim {\mbox{\boldmath$s$%
\unboldmath}}\cdot ({\mbox{\boldmath$\pi$\unboldmath}}\times \mathbf{E}-%
\mathbf{E}\times {\mbox{\boldmath$\pi$\unboldmath}})$ describing a
spin-orbit coupling and, finally, a term $\sim \texttt{div}\mathbf{E}$,
i.e., a Darwin coupling.

Similarly, starting with equation (\ref{paul3}), setting $\lambda_2=-1,\
\lambda_1=\frac12$, supposing $\frac1\nu$ to be a small parameter and making
use of the transformation $\psi_1\to\psi^{\prime }_1=\hat U\psi_1$ with $%
\hat U=\exp(-\frac{2\mathrm{i}}{\nu m}\mathbf{s}\cdot {\mbox{\boldmath$\pi$%
\unboldmath}})$ we obtain the equation 
\begin{equation}  \label{orb1}
\begin{array}{c}
\left(\pi_0-\frac{\nu^2}2m- \frac{{\mbox{\boldmath$\pi$\unboldmath}}^2}{2m}
-eA_0- \frac{\lambda e} {m^2} \left({\mbox{\boldmath$s$\unboldmath}}\cdot ({%
\mbox{\boldmath$\pi$\unboldmath}}\times\mathbf{E}-\mathbf{E}\times{%
\mbox{\boldmath$\pi$\unboldmath}})+\frac43\texttt{div}
\mathbf{E}\right.\right.\\\left.\left.
-Q_{ab}\frac{\partial E_a}{\partial x_b}\right)+\cdots\right)\psi^{\prime }_1=0.%
\end{array}%
\end{equation}
Here, $\lambda=\frac{2}{\nu^2}$ and the dots denote small terms of the order $%
o(\frac1{\nu^3})$ and $o(e^2)$.

Like equation (\ref{orb}), the equation (\ref{orb1}) includes the terms
which describe the spin-orbit and Darwin couplings. In addition,
there is
 the term $\sim Q_{ab}\frac{\partial E_{a}}{\partial x_{b}}$ which
describes a quadrupole interaction of a charged vector particle with an
electric field.

Let us stress that these terms are kept for the minimal coupling
also, i.e., when we set $\lambda_1=\lambda_2=0$ in (\ref{anint1})
and (\ref{Ha}).

Analogously we can analyze equations (\ref{paul3}) with the Hamiltonias (\ref%
{P24}) and (\ref{P25}). Setting there $g=0$ and making transformations $\hat{%
H}\rightarrow U\hat{H}U^{-1}-iU\frac{\partial U^{-1}}{\partial t},$ where $%
U=\exp (-3\mathrm{i}\mathbf{K}\cdot {\mbox{\boldmath$\pi$\unboldmath}}/2m)$
and $U=\exp (-3\mathrm{i}\mathbf{\hat{K}}\cdot {\mbox{\boldmath$\pi$%
\unboldmath}}/2m)$ for Hamiltonian (\ref{P24}) and (\ref{P25}),
respectively, we obtain the approximate equation (\ref{orb1}) with $\lambda
=9\nu /8,\ \nu =\mp 1$.

Thus we again come to the conclusion (see \cite{nikitin}) that the
spin-orbit and Darwin couplings can be effectively described within the
framework of a Galilei-invariant approach and thus they have not be
necessarily interpreted as purely relativistic effects.

Let us note that it is possible to choose parameters $\lambda_1$ and $%
\lambda_2$ in (\ref{Ha}) in such a way that the anomalous interaction with
an electric field will not be present. Namely, we can set $\lambda_2=1,
\lambda_1=-1/2$, and obtain, instead of (\ref{orb1}), the following equation
:
\begin{equation}  \label{orb2}
\begin{array}{l}
\left(\pi_0-\frac{\nu^2}2m- \frac{{\mbox{\boldmath$\pi$\unboldmath}}^2}{2m}
-eA_0+\frac{ge} {2m}\mathbf{s}\cdot\mathbf{H} +\cdots\right)\psi^{\prime
}_1=0,%
\end{array}%
\end{equation}
where $g=2$. In other words, introducing a specific anomalous interaction
into the Galilean Duffin-Kemmer equation we can reduce it to the
Schr\"odinger-Pauli equation with the correct value of gyromagnetic ratio $g$%
.

\section{\label{S5}Discussion}

It is pretty well known that the correct definition of non-relativistic
limit of relativistic theories is by no means a simple problem. In
particular, as it was noticed once more in the recent paper \cite{New}, a
straightforward non-relativistic expansion in terms of $v/c$ leads to
loosing either Galilei-invariance or important contributions such as
spin-orbit coupling. Thus to make this limit correctly it is imperative to
have \textit{a priory} information on possible Galilean limits of a given
theory.

In the present paper we continue the study of the Galilei-invariant
theories for vector and spinor fields, started in
\cite{Marc}-\cite{NN3}. The peculiarity of our
approach is that, as distinct to the other approaches (e.g., to \cite{ll1967}%
-\cite{ourletter}, \cite{santos} ), it enables to find out {\it a
complete list of the Galilei-invariant equations for massive scalar
and vector fields}. This possibility is due to our knowledge of all
non-equivalent indecomposable representations of the Galilei algebra
$hg(1,3)$ that can be constructed on representation spaces of scalar
and vector fields, i.e., the representations which
 were described
for the first time in paper \cite{Marc}. Notice that Galilean
equations for massless vector and scalar fields are presented in our
previous paper \cite{NN3}.

Using this complete list of representations we find all systems of the first
order Galilei-invariant wave equations (\ref{equ}) for scalar and vector
fields. The $\beta $-matrices for these Galilei-invariant wave equations are
given in the Appendix A. In fact we have described how to
construct any wave equation of finite order invariant
with respect to
the Galilei group since such equation can be written as a first order
differential equation in which various derivatives of fields are considered
as new variables of $\Psi.$

Then Galilean analogues of some popular relativistic equations for vector
particles and particles with spin 3/2 are described, in particular, the
Galilean Proca and the Galilean Rarita-Schwinger equations.
However these Galilean
equations are not non-relativistic limits of the relativistic
Proca or relativistic Rarita-Schwinger equations since,
 among other things, they have more
components. Thanks to that it is possible to obtain equations which keep all
the main features of their relativistic analogues. To the best of our
knowledge this is done for the first time in the present paper.

We pay a special attention to description of the Galilean particles
interacting with an external electromagnetic field. We study
both the cases
 with a
minimal interaction as well as anomalous one.

A quite general form of an anomalous interaction which satisfies the Galilei
invariance condition is written in equation (\ref{anint1}). It contains two coupling
constants, $\lambda _{1}$ and $\lambda _{2}$, whose values can be fixed via
physical reasonings. If we fix the value of gyromagnetic ratio predicted by (%
\ref{anint1}) then the randomness in description of anomalous interaction is
reduced to one arbitrary parameter. It can be fixed too if we restrict
ourselves to a desired value of the spin-orbit coupling constant.

Notice that the results presented in Sections 4.1 and 4.3 are
valid for
arbitrary equations (\ref{anint1}) invariant with respect to the Galilei
group whereas those in Sections 4.2, 4.4 and 4.5 are true for
special
equations with anomalous interactions, i.e.,
for the Galilean Proca equation,
the generalized L\'{e}vy-Leblond
equation and for a generalized Galilean
Duffin-Kemmer equation.
We shall show that the last equations describe
consistently charged particles interacting with an
 electromagnetic
field. In other words, they describe an important physical effect,
namely,
spin-orbit coupling which is, however, traditionally interpreted as a purely
relativistic phenomenon.

On the other hand let us note that there are some principal
difficulties with the Galilean approach since the
Galilei invariance requires that mass and
energy are separately conserved, and that within the Galilean
theories there
is not concept of proper time which yields a phase effects that does
 not
depend on the velocity of light and so does not disappear in a
non-relativistic limit. Of course, there are obvious restrictions to
phenomena which are characterized by velocities much smaller than that
 of light. Moreover,  there are also problems in our approach with
interpretation of undesired terms $\sim Q_{ab}\frac{\partial H_{a}}{\partial
x_{b}}$ and $\mathbf{s}\cdot ({\mbox{\boldmath$\pi$\unboldmath}}\times
\mathbf{H}-\mathbf{H}\times {\mbox{\boldmath$\pi$\unboldmath}})$ which
appear in Hamiltonian (\ref{paul3}). Thanks to an appropriate
choice of otherwise
arbitrary parameters $\lambda _{1}$ and $\lambda _{2}$ these terms are not
present in effective Hamiltonians (\ref{orb1}) as well as  (\ref{orb2}) which
describe spin-orbit and the Pauli couplings respectively.
However, if we
would like to keep both these couplings, then the undesired terms
may appear.

Another problem is connected with a sign in front of the terms describing
the spin-orbit and the Darwin couplings. Comparing (\ref{orb}) with the
quasi-relativistic approximation of the Dirac equation we conclude,
that in order to
obtain a correct signs it is necessary to suppose that
$\lambda _{3}$ be pure imaginary. Moreover, for $\lambda _{3}=\mathrm{i}$ the coupling
constants for spin-orbit and the Darwin interactions in (\ref{orb}) coincide
with the relativistic ones predicted by the Dirac equation.

Notice that the exact equation (\ref{paulik})
is much simpler then the
approximate equation (\ref{orb}) and can be solved exactly for some
particular external fields (for instance, the Coulomb ones).
However, if $\lambda _{3}$ is
imaginary, the term $-\frac{e\lambda _{3}}{2m}{\mbox{\boldmath$\sigma$%
\unboldmath}}\cdot \mathbf{F}$ in equation (\ref{paulik}) and the
corresponding Hamiltonian $\hat{H}=-L+p_{0}=A_{0}$
are non-hermitian. If $A_{0}$ and $\mathbf{A}$ are even and odd
functions of $%
\mathbf{x,}$ respectively, the equation (\ref{paulik}) appears to be
invariant with respect to a product of the space inversion $P$ and the
Wigner time inversion $T$ (compare with \cite{bender}), and thus serves as
an example of a $PT$ - symmetric quantum-mechanical system.

Any Galilean theory, by definition, is only an approximation of the corresponding
relativistic one. Thus the very existence of a physically consistent
non-relativistic approximation can serve as a consistency criterium of a
relativistic theory. Thus our study of the Galilean wave equations
has
contributed
to the theory of relativistic equations,
since effectively we have
analyzed possible non-relativistic limits of theories
for vector and scalar
particles.  On the other hand the results of the present paper can be
applied to purely non-relativistic models satisfying the Galilei
invariance criteria or being invariant w.r.t. various extensions of the
Galilei group, e.g., w.r.t. Galilei supergroup \cite{han}.

\renewcommand{\theequation}{A\arabic{equation}} %
\setcounter{equation}{0} \appendix

\section{Submatrices $R$ and $E$ of matrices $\beta_4$}

Here we present all non-trivial solutions of equations (\ref{b2})
which give rise to explicit forms of matrices $\beta_4$ given
by equation (\ref{b1}). The associated matrices $\beta_0$ and $\beta_a$
are given by equations (\ref{b3}) and (\ref{b4}).

Solving equations (\ref{b2}), where $A$, $C$ and $A'$, $C'$ are
matrices given in Table 1 which correspond to $q=(n,m,\lambda)$ and
$q'=(n', m', \lambda')$ respectively we obtain the associated
matrices $R=R(q,q'), E= E(q,q')$ (which determine matrix $\beta_4$
via (\ref{b1})) in the forms presented in Tables 2-4, where the
Greek letters denote arbitrary real parameters.
\newpage
\begin{center}

Table 2. Submatrices $R$ and $E$ of matrices $\beta_4$

\end{center}
\begin{tabular}{|c|c|c|c|}
\hline
& \multicolumn{3}{|c|}{$m,n,\lambda$}\\
\cline{2-4} \raisebox{1.8ex}[0pt][0pt]{$m',n',\lambda'$}
&3,1,1&2,2,1&
2,1,0\\
\hline
3,1,1&$\bea{l}R=\left(\bea{ccc}\mu&\nu&\sigma\\\nu&\alpha&\lambda\\
\sigma&\lambda&0\eea\right)
\\
E=\alpha-2\sigma,\\ \mu\nu=0, \lambda(\alpha-\sigma)=0\eea$&
$\bea{c}R=\left(\bea{ccc}\mu&\sigma&\omega\\\nu&\alpha&0\eea\right)\\
E=\left(\bea{c}\kappa\\\omega-\alpha\eea\right)\eea$&$\bea{c}
R=\left(\bea{ccc}\mu&\sigma&0\\\nu&\alpha&\omega\eea\right)\\\\
E=\bea{c}\kappa\eea\eea$\\\hline
2,2,1&$\bea{l}R=\left(\bea{cc}\mu&\nu\\\sigma&\alpha\\\omega&0\eea\right)\\
E=(\kappa\ \ (\omega-\alpha))\eea$&$\bea{l}R=\left(\bea{cc}\mu&\nu\\
\nu&\kappa\eea\right),\ \mu\nu=0\\
E=\left(\bea{cc}\sigma&0\\0&\omega\eea\right)\eea$&$\bea{l}R=\left(
\bea{cc}\mu&\sigma\\\nu&\omega\eea\right)\\\\E=(\kappa\
\omega)\eea$\\\hline
2,1,0&$\bea{l}R=\left(\bea{ll}\mu&\nu\\\sigma&\alpha\\0&\omega\eea\right)\\
E=\kappa\eea$&$\bea{l}R=\left(\bea{cc}\mu&\nu\\
\sigma&\omega\eea\right),E=\left(\bea{c}\kappa\\\omega\eea\right)\eea$&$\bea{l}
R=\left(\bea{cc}\mu&\nu\\\nu&\kappa\eea\right)\\
E=\sigma,\ \mu\nu=0\eea$\\\hline
2,1,1&$\bea{l}R=\left(\bea{cc}\mu&\nu\\\sigma&\alpha\\\omega&0\eea\right)\\
E=\omega-\alpha\eea$&$\bea{l}R=\left(\bea{cc}\mu&\nu\\
0&\omega\eea\right),
E=\left(\bea{c}\alpha\\\sigma\eea\right)\eea$&$\bea{l}R=\left(\bea{cc}
\mu&\sigma\\0&\nu\eea\right)\\ E=\kappa\eea$\\\hline
2,0,0&$\bea{l}R=\left(\bea{cc}\mu&\nu\\\sigma&\alpha\\\alpha&0\eea\right)\\
E
\texttt{ not existing}\eea$&$\bea{l}R=\left(\bea{cc}\mu&\nu\\
\omega&0\eea\right)\\ E\ \texttt{not
existing}\eea$&$\bea{l}R=\left(\bea{cc}\mu&\nu\\\sigma&0\eea\right)\\
E\ \texttt{not}\\ \texttt{existing}\eea$\\\hline 1,2,1&$\bea{l}
R=\left(\bea{c}\mu\\\nu\\\alpha\eea\right)\\E= (\omega\
\alpha)\eea$&$\bea{l}R=\left(\bea{c}\kappa\\\sigma\eea\right),
E=\left(\bea{cc}\mu&\nu\\
\omega&0\eea\right)\eea$&$\bea{l}R=\left(\bea{c}\mu\\\nu\eea\right)\\
E=(\sigma \ \ 0)\eea$\\\hline
1,1,0&$\bea{l}R=\left(\bea{c}\mu\\\nu\\\alpha\eea\right),  E=
\alpha\eea$&$\bea{l}R=\left(\bea{c}\kappa\\\sigma\eea\right),
E=\left(\bea{c}\mu\\
0\eea\right)\eea$&$\bea{l}  R=\left(\bea{c}\mu\\\nu\eea\right)\\
E=\sigma\eea$\\\hline
1,1,1&$\bea{l}R=\left(\bea{c}0\\\nu\\\alpha\eea\right),\  E=
\omega\eea$&$\bea{l}R=\left(\bea{c}\kappa\\\sigma\eea\right),
E=\left(\bea{c}\mu\\
\nu\eea\right)\eea$&$\bea{l}R=\left(\bea{c}\mu\\\nu\eea\right)\\
E=0\eea$\\\hline
1,0,0&$\bea{l}R=\left(\bea{c}\mu\\\alpha\\0\eea\right)\\  E
\texttt{ not existing}\eea$&$\bea{l}R=\left(\bea{c}\kappa\\\sigma\eea\right)\\
E\ \texttt{not existing}\eea$&$\bea{l}R=\left(\bea{c}\kappa\\\sigma\eea\right)\\
E\ \texttt{not}\\ \texttt{existing}\eea$\\\hline 0,1,0&$\bea{l}R\
\texttt{not existing},\\  E=\alpha\eea$&$\bea{l}R\ \texttt{ not}\\
\texttt{existing},\eea E=\left(\bea{c}\kappa\\\sigma\eea\right)
$&$\bea{l}E=\alpha,\  R\ \texttt{not}\\\texttt{existing}
\eea$\\\hline

\end{tabular}

\newpage

\begin{center}
Table 3. Submatrices $R$ and $E$ of matrices $\beta_4$
\end{center}
 \begin{tabular}{|c|c|c|c|}
\hline
& \multicolumn{3}{|c|}{$m,n,\lambda$}\\
\cline{2-4} \raisebox{1.8ex}[0pt][0pt]{$m',n',\lambda'$}
&2,1,1&2,0,0&1,2,1
\\
\hline
2,1,1&$\bea{l}R=\left(\bea{cc}\mu&\nu\\\nu&0\eea\right)\\
E=\sigma,\ \mu\nu=0\eea$&$\bea{l}R=\left(\bea{cc}\omega&\nu\\
\mu&0\eea\right)\\E\ \texttt{not}
\ \texttt{existing}\eea$&$\bea{l}R=(\mu\ \nu)\\
E=\left(\bea{c}\sigma\\\alpha\eea\right)\eea$\\\hline
2,0,0&$\bea{l}R=\left(\bea{cc}\omega&\mu\\
\nu&0\eea\right)\\E\ \texttt{not}
\ \texttt{existing}\eea$&$\bea{l}R=\left(\bea{cc}\mu&\nu\\
\nu&0\eea\right),
E\ \texttt{not} \\ \texttt{
existing},\ \mu\nu=0\eea$&$\bea{l}R=\left(\bea{cc}\mu&\nu\eea\right)\\
E\ \texttt{not}  \\\texttt{existing}\eea$\\\hline
1,2,1&$\bea{l}R=\left(\bea{c}\mu\\\nu\eea\right)\\
E=(\sigma \ \ \alpha)\eea$&$\bea{l}R=\left(\bea{c}\mu\\\nu\eea\right)\\
E\ \texttt{not}  \ \texttt{existing}\eea$&$\bea{l}E=\left(\bea{cc}\mu&\nu\\
\nu&0\eea\right),
\\\mu\nu=0; R=\alpha
\eea$\\\hline 1,1,0&$\bea{l}R=\left(\bea{c}\mu\\\nu\eea\right),\
E=\sigma\eea$&$\bea{l}R=\mu,\ \
 E\ \texttt{not} \\\texttt{
existing}\eea$&$\bea{l}R=\mu\\
E=\left(\bea{c}\nu\\0\eea\right)\eea$\\\hline
1,1,1&$\bea{l}R=\left(\bea{c}\mu\\\nu\eea\right), \
E=\sigma\eea$&$\bea{l}R=\mu,\ \
 E\ \texttt{not} \\\texttt{
existing}\eea$&$\bea{l} R=\mu,\
E=\left(\bea{c}\nu\\\alpha\eea\right)\eea$\\\hline
1,0,0&$\bea{l}R=\left(\bea{c}\kappa\\\sigma\eea\right)\\
E\ \texttt{not} \ \texttt{ existing}\eea$&$\bea{l}R=\mu,\ \
 E\ \texttt{not} \\\texttt{
existing}\eea$&$\bea{l}R=\mu,\\
 E\ \texttt{not} \texttt{
existing}\eea$\\\hline 0,1,0&$\bea{l}R\ \texttt{not existing},\\
E=\alpha\eea $&$\bea{l}R \texttt{ and }E\\ \texttt{not
existing}\eea$&$\bea{l} R \ \texttt{not existing}\\
E=\mu\eea$\\\hline

\end{tabular}


\begin{center}

Table 4. Submatrices $R$ and $E$ of matrices $\beta_4$
\end{center}
 \begin{tabular}{|c|c|c|c|c|}
\hline
& \multicolumn{4}{|c|}{$m,n,\lambda$}\\
\cline{2-5} \raisebox{1.8ex}[0pt][0pt]{$m',n',\lambda'$}
&1,1,0&1,1,1&1,0,0&0,1,0
\\
\hline 1,1,0&$\bea{l}R=\mu\\ E=\nu\eea$&$\bea{l}R=\mu\\
E=\nu\eea$&$\bea{l}
 R=\mu, E\ \texttt{not} \\\texttt{existing} \\\eea$&$\bea{l}E=\mu,
 R\ \texttt{not} \\\texttt{existing} \eea$\\\hline 1,1,1&$\bea{l}R=\mu,
 \\ E=\nu\eea$&$\bea{l}R=\mu\\
E=0 \eea$&$\bea{l}R=\mu,
 E\ \texttt{not} \\\texttt{
existing} \eea$&$\bea{l}E=\mu,
 R\ \texttt{not} \\\texttt{
existing} \eea$\\\hline 1,0,0&$\bea{l}R=\mu,
 E\ \texttt{not} \\\texttt{
existing} \eea$&$\bea{l}R=\mu,
 E\ \texttt{not} \\\texttt{existing} \eea$&$\bea{l}R=\mu, E\texttt{ not}
 \\\texttt{existing}
\eea$&$\bea{l}R \texttt{ and }
 E\\ \texttt{not
existing} \eea$\\\hline 0,1,0&$\bea{l}E=\mu,
 R\ \texttt{not} \\\texttt{
existing} \eea$&$\bea{l}E=\mu,
 R\ \texttt{not} \\\texttt{
existing} \eea$&$\bea{l}R \ \texttt{and }E\\\texttt{not
existing}\eea$&$\bea{l}E=\mu,R \ \texttt{not} \\ \texttt{existing}
\eea$\\\hline
\end{tabular}

\vspace{2mm}

These matrices are simplified using the equivalence transformations
\beq\label{llast}R\to URU^{\dag}$, $E\to VEV^{\dag}.\eeq Here $U$
and $V$ are unitary matrices whose dimensions are the same as
dimensions of matrices $R$ and $E$ correspondingly, which satisfy
the following relations: $UA=AU,\ UB=BV,\ VC=CU$. Transformations
(\ref{llast}) keep equations (\ref{b2}) invariant.

\renewcommand{\theequation}{B\arabic{equation}} %
\setcounter{equation}{0}

\section{More on the Galilean Rarita-Schwinger equation}
Let us prove that equation (\ref{ra}) is consistent and describes a particle with
spin $s=3/2$.

Reducing (\ref{ra}) by $p_\texttt{m}$ we obtain $\lambda m^2\hat\Psi^4=0$,
i.e., $\hat\Psi^4=0$. Whereas, reducing (\ref{ra})
by $\hat\gamma_\texttt{m}$
we obtain
\begin{equation}  \label{con}
p_\texttt{n}\hat\Psi^\texttt{n}= \hat\gamma_\texttt{n} p^\texttt{n}%
\hat\gamma_\texttt{m}\hat \Psi^\texttt{m}.
\end{equation}
Finally, comparing (\ref{con}) with equation (\ref{ra})
for index $\texttt{m}=4$
we find the following consequences of equation (\ref{ra}):
\begin{gather}  \label{ra1}
\hat\gamma_\texttt{n} p^\texttt{n}\hat\Psi^\sigma=0,\ \ \ \ \sigma=0,1,2,3,
\\\label{ra2}
m\hat\Psi^0-p^a\hat\Psi^a=0,\ \ \ a=1,2,3, \\\label{ra3}
\hat\gamma_0\hat\Psi^0+\hat\gamma_a\hat\Psi^a=0,\ \texttt{and } \hat\Psi^4=0.
\end{gather}
On the other hand equation (\ref{ra}) follows from (\ref{ra1})--(\ref{ra3}),
so that equations (\ref{ra}) and (\ref{ra1})--(\ref{ra3}) are equivalent.

In accordance with (\ref{ra1}) any component of $\Psi^\texttt{m}$ satisfies
the L\'evy-Leblond equation (compare with Section \ref{S21}).
Let us prove now
that equations (\ref{ra1})--(\ref{ra3}) describe indeed a particle with spin
s=3/2.

It follows from (\ref{ra1})--(\ref{ra3}) that, in the rest frame, $%
\hat\Psi^0=\hat\Psi^4=0 $ and $\hat\Psi^a$ has only two non-zero
spinor
components $\hat\Psi^a_\alpha, \ \alpha=1,2$. Using
$\hat\gamma$-matrices in realization
(\ref%
{gamma}) and equation (\ref%
{ra3}), we conclude that $\hat\Psi^a$ satisfies the equation
\begin{equation}  \label{?}
\sigma_a\Psi^a=0,.
\end{equation}
Consequently this function satisfies conditions (\ref{cas6}) and
(\ref{cas61}) as well
with $s=3/2$. This follows from the fact the total spin
operator $\mathbf{S}$ is a sum of
operators of spin one and of spin one--half: ${\ S_a}= s_a+\frac12 \sigma_a$%
, so that that
\begin{equation}  \label{rs12}
\mathbf{S}^2=\frac{11}{4}+\mathbf{s}\cdot{\mbox{\boldmath$\sigma$\unboldmath}%
}.
\end{equation}

Let $\tilde\Psi$ denotes the column $(\hat\Psi^1, \hat\Psi^2, \hat\Psi^3)$.
In accordance with (\ref{rs12}) the condition $\mathbf{S}^2\tilde\Psi=s(s+1)%
\tilde\Psi$ reduces to the form
\begin{equation}  \label{rs13}
\hat\Psi_a-\frac{i}{2}\varepsilon_{abc} \sigma_b\hat\Psi_c=0,
\end{equation}
for $s=3/2$ provided we use the representation with $(s_a)_{bc}=i%
\varepsilon_{abc} $, where $\varepsilon_{abc}$ is a totally antisymmetric
unit tensor.

Comparing (\ref{?}) with (\ref{rs13}) we conclude that these equations are
completely equivalent since multiplying (\ref{?}) by $\hat\sigma_a$ we
obtain (\ref{rs13}) and multiplying (\ref{rs13}) by $\hat s_a$ and
contracting it with respect to index $a$ we get (\ref{?}).

Thus indeed equation (\ref{ra}) describes a Galilean particle with spin $%
s=3/2$.
\renewcommand{\theequation}{C\arabic{equation}} %
\setcounter{equation}{0}
\section{Contraction of relativistic Proca equation}

Let us consider the relativistic Proca equation (\ref{001}) and contract it
directly to\ a non-relativistic (i.e., Galilei-invariant) approximation.
Solutions of this equation are a four-vector $\Psi ^{\mu }$ and a
skew-symmetric tensor $\Psi ^{\mu \nu }$ which transform according to the
representation $D(\frac{1}{2},\frac{1}{2})\oplus D(1,0)\oplus D(0,1)$ of the
Lorentz group.

It was shown in papers \cite{Marc} and \cite{NN2} how this representation
can be contracted to representation $D(3,1,1)$ of the homogeneous Galilei
group. This contraction can be used to reduce equation (\ref{001}) to a
Galilei-invariant form. To do this it is necessary:

\begin{itemize}
\item To choose the following new dependent variables
\begin{equation*}
\begin{array}{l}
R^a=-\frac12(\Psi^{0a}+\Psi^a),\ \ N^a=\Psi^{0a}-\Psi^a, \\
W^c=\frac12\varepsilon^{abc}\Psi_{bc}, \ \ B=\Psi^0,%
\end{array}%
\end{equation*}
which, in accordance with (\ref{001}), satisfy the following equations:
\begin{equation}  \label{eququ}
\begin{array}{l}
2(p^0-\kappa )R^a+p^aB+\varepsilon^{abc}p_bW_c=0, \\
(p^0+\kappa )N^a-\varepsilon^{abc}p_bW_c+p^aB=0, \\
\varepsilon^{abc}p_b(R_c+ \frac12N_c)=\kappa W^a, \\
\frac12p_aN^a-p_aR^a=\kappa B;%
\end{array}%
\end{equation}

\item To act on variables $R^{a},N^{a},W^{a}$ and $B$ by a diagonal
contraction matrix. This action yields the change:
\begin{equation*}
R^{a}=\tilde{R}^{a},\ N^{a}=\varepsilon ^{2}\tilde{N}^{a},\
W^{a}=\varepsilon \tilde{W}^{a},\ B=\varepsilon \tilde{B},
\end{equation*}%
where $\varepsilon $ is a small parameter associated with the inverse speed
of light;

\item To change relativistic four-momentum $p^{\mu }$ and mass $\kappa $ by
their Galilean counterparts $\tilde{p}^{a},\tilde{p}^{0}$ and $m,$ where
\begin{equation*}
\text{ \ \ }\tilde{p}^{a}=\varepsilon ^{-1}p^{a},\ \tilde{p}%
^{0}=p_{0}-\kappa \texttt{\ and  }m=\frac{1}{2}(p_{0}+\kappa )\varepsilon
^{-2};
\end{equation*}

\item In each equation in (\ref{eququ}) keep only terms which are multiplied
by the lowest powers of $\varepsilon $.
\end{itemize}

As a result we obtain the following equations for $\tilde R^a, \tilde N^a,
\tilde W^a$ and $\tilde B$:
\begin{equation}  \label{equq}
\begin{array}{l}
2\tilde p^0 \tilde R^a+\tilde p^a\tilde B+ \varepsilon^{abc}\tilde p_b\tilde
W_c=0, \\
\varepsilon^{abc}\tilde p_b\tilde R_c =m\tilde W^a, \\
\tilde p_a\tilde R^a+m\tilde B=0,%
\end{array}%
\end{equation}
and
\begin{equation}  \label{aquq}
2m\tilde N^a=\varepsilon^{abc}\tilde p_b\tilde W_c- \tilde p^a\tilde B.
\end{equation}

The system (\ref{equq}) is nothing else but the Galilei-invariant equation (%
\ref{equ}) with matrices (\ref{M2}) written componentwise. Relation (\ref%
{aquq}) expresses the extra component $\tilde N^a$ via derivatives of the
essential ones, i.e. of $\tilde W_c$ and $\tilde B$.

Thus the Galilean analogue of the Proca equation (\ref{P12}) cannot be
obtained as a non-relativistic limit of the relativistic Proca equation (\ref%
{001}) but is a specific modification of it. The relativistic counterpart of
equation (\ref{P12}) is a specific generalization of (\ref{001}) which will
be studied in a separate publication.

\end{document}